\pdfoutput=1

\documentclass[11pt,twoside,a4paper,cmspaper,final,collab]{cms-tdr}
\def\svnVersion{415782}\def\cmsCernNoTag{CERN-EP/2017-031}\def\cmsCernDate{\today}\def\cmsMessage{Published in the Journal of High Energy Physics as \href{http://dx.doi.org/10.1007/JHEP07(2017)014}{\doi{10.1007/JHEP07(2017)014.}}}

\begin{document}\cmsNoteHeader{EXO-16-037}

\hyphenation{had-ron-i-za-tion}
\hyphenation{cal-or-i-me-ter}
\hyphenation{de-vices}
\RCS$HeadURL: svn+ssh://svn.cern.ch/reps/tdr2/papers/EXO-16-037/trunk/EXO-16-037.tex $
\RCS$Id: EXO-16-037.tex 414379 2017-07-03 18:59:07Z avartak $
\newlength\cmsFigWidth

\ifthenelse{\boolean{cms@external}}{\setlength\cmsFigWidth{0.85\columnwidth}}{\setlength\cmsFigWidth{0.4\textwidth}}
\ifthenelse{\boolean{cms@external}}{\providecommand{\cmsLeft}{top\xspace}}{\providecommand{\cmsLeft}{left\xspace}}
\ifthenelse{\boolean{cms@external}}{\providecommand{\cmsRight}{bottom\xspace}}{\providecommand{\cmsRight}{right\xspace}}

\newcommand{\muz}{\ensuremath{\boldsymbol{\mu}^{\Zvv}}}
\providecommand{\mt}{\ensuremath{m_{\mathrm{T}}\xspace}}
\newcommand{\Zmm}{\ensuremath{\Z(\PGmp\PGmm)}}
\newcommand{\Zee}{\ensuremath{\Z(\Pep\Pem)}}
\newcommand{\Zll}{\ensuremath{\Z(\ell^{+} \ell^{-})}}
\newcommand{\Zvv}{\ensuremath{\Z(\PGn\PAGn})}
\newcommand{\Wmn}{\ensuremath{\PW(\mu\nu)}}
\newcommand{\Wen}{\ensuremath{\PW(\Pe\nu)}}
\newcommand{\Zlljets}{\ensuremath{\Z(\ell^{+} \ell^{-})}+\text{jets}}
\newcommand{\phojets}{\ensuremath{\gamma}+jets}
\newcommand{\Zjets}{\Z{}+jets}
\newcommand{\Wjets}{\PW{}+jets}
\newcommand{\Zvvjets}{\ensuremath{\Z(\PGn\PAGn)}+jets}
\newcommand{\Wlvjets}{\ensuremath{\PW(\ell\nu)}+jets}
\newcommand{\brhinv}{\ensuremath{\mathcal{B}(\PH\to\text{inv})}}
\newcommand{\hinv}{\ensuremath{\PH\to \text{inv}}}
\newcommand{\mettrig}{\ensuremath{E_{\mathrm{T, trig}}^{\text{miss}}}}
\newcommand{\mhttrig}{\ensuremath{H_{\mathrm{T, trig}}^{\text{miss}}}}

\newcommand{\x}{\ensuremath{\phantom{0}}}
\cmsNoteHeader{EXO-16-037}
\title{Search for dark matter produced with an energetic jet or a hadronically decaying W or Z boson at $\sqrt{s} = 13\TeV$}

\date{\today}

\abstract{
A search for dark matter particles is performed using events with large missing transverse momentum, at least one energetic jet, and no leptons, in
proton-proton collisions at $\sqrt{s} = 13\TeV$ collected with the CMS detector at the LHC. The data sample
corresponds to an integrated luminosity of 12.9\fbinv. The search includes events with jets from the hadronic decays of a W or Z boson.
The data are found to be in agreement with the predicted background contributions from standard model processes.
The results are presented in terms of simplified models in which dark matter particles are produced through interactions involving a vector, axial-vector, scalar, or pseudoscalar mediator.
Vector and axial-vector mediator particles with masses up to 1.95\TeV, and scalar and pseudoscalar mediator particles with masses up to 100 and 430\GeV respectively, are excluded at
95\% confidence level. The results are also interpreted in terms of the invisible decays of the Higgs boson, yielding an observed (expected)
95\% confidence level upper limit of 0.44 (0.56) on the corresponding branching fraction.
The results of this search provide the strongest constraints on the dark matter pair production cross section through vector and axial-vector mediators at a particle collider.
When compared to the direct detection experiments, the limits obtained from this search provide stronger constraints for dark matter masses less than 5, 9, and 550\GeV,
assuming vector, scalar, and axial-vector mediators, respectively. The search yields stronger constraints for dark matter masses less than 200\GeV, assuming
a pseudoscalar mediator, when compared to the indirect detection results from Fermi--LAT.
}

\hypersetup{%
pdfauthor={CMS Collaboration},%
pdftitle={Search for dark matter with an energetic jet or a hadronically decaying W or Z boson at sqrt(s) = 13 TeV},%
pdfsubject={CMS},%
pdfkeywords={CMS, physics, exotica, dark matter, monojet}}

\maketitle

\section{Introduction}

Astrophysical observations have provided compelling evidence for the existence of dark matter (DM) in the universe~\cite{Bertone:2004pz,Feng:2010gw,Porter:2011nv}.
However, there is no compelling experimental evidence for non-gravitational interactions between the DM and standard model (SM) particles.
Most current models of DM assume that it consists of weakly interacting massive particles (WIMPs)~\cite{Feng:2010gw}. If such particles exist,
direct pair production of WIMPs may occur in TeV-scale collisions at the CERN LHC~\cite{Beltran:2010ww}.
If DM particles are produced at the LHC, they would not generate directly observable signals in the detector. However, if they recoil
against a jet radiated from the initial state, they may produce an apparent, large transverse momentum imbalance in the event.
This is termed the `monojet' final state~\cite{Goodman:2010ku,Bai:2010hh}. The DM particles may also be produced in association with an electroweak boson,
resulting in the `mono-V' signature, where V represents the W or Z boson~\cite{Bell:2012rg,An:2012ue,Carpenter:2012rg}.
Observation of these final states could be interpreted as evidence for DM particles.
Additionally, the Higgs boson~\cite{Aad:2012tfa,Chatrchyan:2012xdj,Chatrchyan:2013lba} could be a mediator between DM
and SM particles~\cite{Lee:2008xy,Baek:2012se,Djouadi:2011aa,Djouadi:2012zc,Beniwal:2015sdl}.
The monojet and mono-V signatures can be used to set a bound on the invisible branching fraction of the Higgs boson.

Several previous searches at the LHC have exploited the mono-V and monojet signatures. Results from earlier searches~\cite{Aad:2013oja,Khachatryan:2014rra,Aad:2015zva}
have typically been interpreted using effective field theories that model contact interactions between the DM and SM particles. Recent search results~\cite{Khachatryan:2016mdm,Aaboud:2016tnv,Aaboud:2016qgg} have been interpreted in terms of simplified DM models~\cite{An:2012va,Buchmueller:2013dya,Malik:2014ggr,Harris:2014hga,Buckley:2014fba,Haisch:2015ioa,Harris:2015kda}.
The invisible branching fraction of the Higgs boson, \brhinv, has been constrained by several searches
at the LHC~\cite{Aad:2014iia,Chatrchyan:2014tja,Aad:2015uga,Aad:2015txa,Aad:2015zva}, with the ATLAS and CMS Collaborations setting upper limits of 0.25 and 0.24,
at 95\% confidence level (CL), respectively, through direct searches~\cite{Aad:2015pla,Khachatryan:2016whc}.
Precise measurements of the Higgs boson couplings from a combination of 7 and 8\TeV data sets, collected by the ATLAS and CMS experiments, provide indirect constraints
on additional contributions to the Higgs boson width from non-SM decay processes. The resulting indirect upper
limit on the Higgs boson branching fraction to non-SM decays is 0.34, at 95\% CL~\cite{Khachatryan:2016vau}.

\begin{figure}[htbp]
\centering
\includegraphics[angle=0,width=0.36\textwidth]{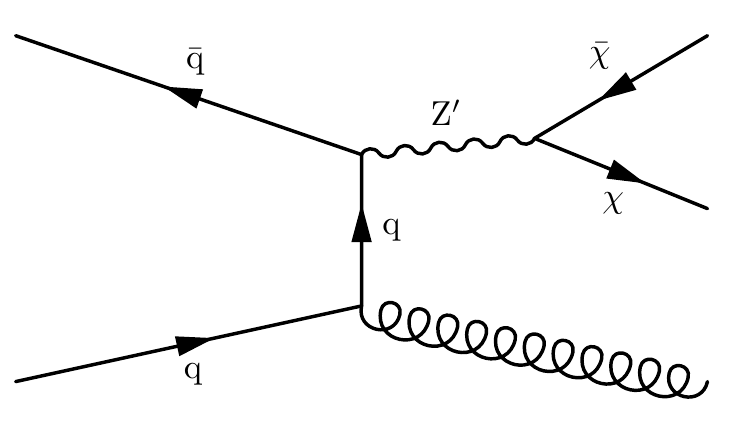}
\includegraphics[angle=0,width=0.36\textwidth]{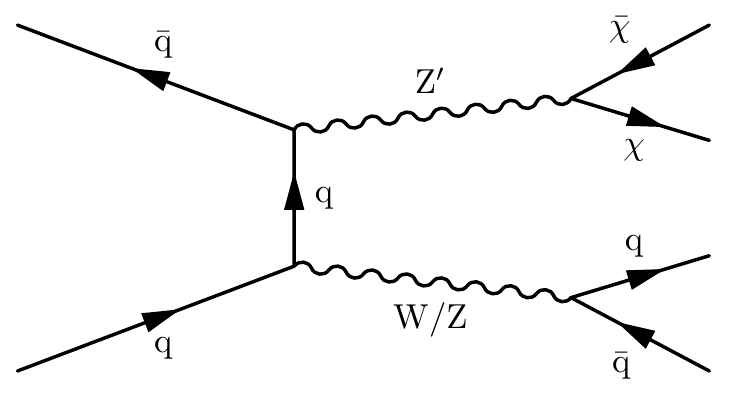}
\caption{Leading order Feynman diagrams of monojet (left) and mono-V (right) production and decay of a spin-1 mediator.}
\label{fig:Spin1_FD}
\end{figure}

\begin{figure}[htbp]
\centering
\includegraphics[angle=0,width=0.36\textwidth]{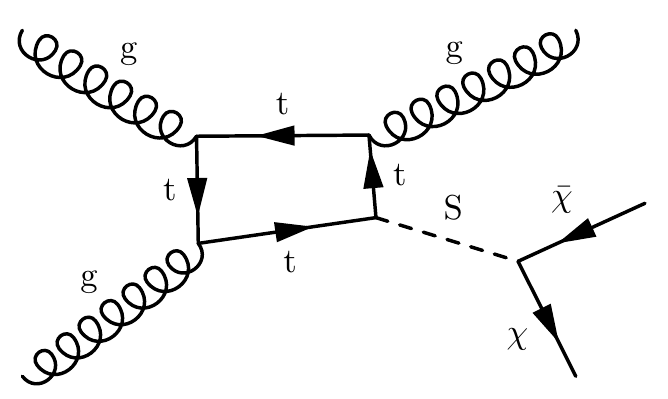}
\includegraphics[angle=0,width=0.36\textwidth]{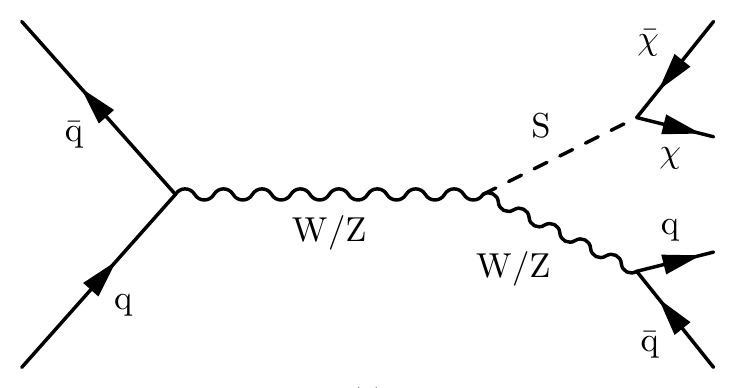}
\caption{Leading order Feynman diagrams of monojet (left) and mono-V (right) production and decay of a spin-0 mediator.}
\label{fig:Spin0_FD}
\end{figure}

This paper presents the results of a search for DM in the mono-V and monojet channels
using a data set of proton-proton collisions at $\sqrt{s} = 13\TeV$, collected with the CMS detector
in the first half of 2016, and corresponding to an integrated luminosity of 12.9\fbinv.
In the case of the mono-V signature, a hadronic decay of a W or Z boson reconstructed as a single large-radius jet is considered.
The results of the search are interpreted using simplified DM models in which the interaction between the DM and SM particles is
mediated by a spin-1 particle such as a $\Z'$ boson, as shown in Fig.~\ref{fig:Spin1_FD},
or a spin-0 particle (S), as shown in Fig.~\ref{fig:Spin0_FD}. The results are also interpreted in terms of \brhinv.
The Feynman diagrams for the production of the SM Higgs boson and its decay to invisible particles resulting in the monojet and mono-V final
states are similar to those shown for a spin-0 mediator in Fig.~\ref{fig:Spin0_FD}.

\section{The CMS detector}
The CMS detector is a multi-purpose apparatus designed to study a wide range of physics processes in proton-proton and heavy ion collisions.
Its central feature is a superconducting solenoid of 6\unit{m} internal diameter that produces a magnetic field of 3.8\unit{T} parallel to the beam direction.
A silicon pixel and strip tracker is contained inside the solenoid and measures the momentum of charged particles up to a pseudorapidity of $|\eta| = 2.5$.
The tracker is surrounded by a lead tungstate crystal electromagnetic calorimeter (ECAL) and a sampling hadron calorimeter (HCAL) made of brass and scintillator,
which provide coverage up to $|\eta| = 3$. The steel and quartz-fiber \v{C}erenkov hadron forward calorimeter extends the coverage to $|\eta| = 5$.
The muon system consists of gas-ionization detectors embedded in the steel flux-return yoke of the solenoid, and covers $|\eta| < 2.4$.
A more detailed description of the CMS detector, together with a definition of the coordinate system used and the relevant kinematic variables,
can be found in Ref.~\cite{Chatrchyan:2008zzk}.

The particle-flow (PF) event algorithm~\cite{CMS-PAS-PFT-09-001,CMS-PAS-PFT-10-001} reconstructs and identifies each individual particle with an optimized combination of information
from the various elements of the CMS detector. The energy of photons is directly obtained from the ECAL measurement.
The energy of electrons is determined from a combination of the electron momentum at the primary interaction vertex as determined by the tracker,
the energy of the corresponding ECAL cluster, and the energy sum of all bremsstrahlung photons spatially compatible with originating from the electron track.
The energy of muons is obtained from the curvature of the corresponding track. The energy of charged hadrons is determined from a combination of their
momentum measured in the tracker and the matching ECAL and HCAL energy deposits, corrected for zero-suppression effects and for the response function
of the calorimeters to hadronic showers. Finally, the energy of neutral hadrons is obtained from the corresponding ECAL and HCAL energies.

The missing transverse momentum vector ($\ptvecmiss$) is computed as the negative vector sum of the transverse momenta (\pt) of all the PF candidates in an event,
and its magnitude is denoted as \MET. Jets are reconstructed by clustering PF candidates using the anti-$\kt$ algorithm~\cite{Cacciari:2008gp}.
Jets clustered with distance parameters of 0.4 and 0.8 are referred to as AK4 and AK8 jets, respectively.
The primary vertex with the largest sum of $\pt^{2}$ of the associated tracks is chosen as the vertex corresponding to the hard interaction in an event.
All charged PF candidates originating from any other vertex are ignored during the jet reconstruction. Jet momentum is determined as the vectorial sum of
all particle momenta in the jet, and is found from simulation to be within 5 to 10\% of the true momentum, over the whole \pt spectrum and detector acceptance.
An offset correction is applied to jet energies to take into account the contribution from additional proton-proton interactions within the same or adjacent
bunch crossings (pileup). Jet energy corrections are derived from simulation and are confirmed with in situ measurements of the energy balance in dijet
and $\gamma$+jet events~\cite{Chatrchyan:2011av}. These are also propagated to the \MET calculation~\cite{CMS-PAS-JME-16-004}.

\section{Event simulation}
The Monte Carlo generators used to simulate various signal and background processes are listed in Table~\ref{tab:MC_Generators}.
Simulated samples of background events are produced for the \Zjets~and \phojets~processes at leading order (LO) with up to four partons in the final state, using \MADGRAPH{}5\_a\MCATNLO 2.2.3~\cite{Alwall:2014hca}. This generator is also used to simulate the
\Wlvjets~process at next-to-leading order (NLO), with up to two partons in the final state, and the quantum chromodynamics (QCD) multijet background at LO.
The \ttbar and single top quark background samples are produced using \textsc{Powheg} 2.0 ~\cite{Nason:2004rx,Frixione:2007vw,Alioli:2010xd},
and a set of diboson samples is produced with \textsc{Pythia 8.205}~\cite{Sjostrand:2014zea}. The monojet DM signal is simulated at
NLO for spin-1 mediators, and at LO for spin-0 mediators with the resolved top quark loop calculations carried out using \textsc{Powheg}~\cite{Haisch:2013ata,Haisch:2015ioa}.
The mono-V DM signal samples are produced at LO with the \textsc{JHUGen} 5.2.5 generator ~\cite{Gao:2010qx,Bolognesi:2012mm,Anderson:2013afp}
for the scalar mediator, and with \MADGRAPH{}5\_a\MCATNLO for the spin-1 mediators. Standard model Higgs boson signal events produced through gluon fusion
and vector boson fusion are generated using \textsc{Powheg}, while SM Higgs boson production in association with W or Z bosons is simulated using
the \textsc{JHUGen} generator.

Events produced by the \MADGRAPH{}5\_a\MCATNLO, \textsc{Powheg}, and \textsc{JHUGen} generators are further processed with \textsc{Pythia}
using the CUETP8M1 tune~\cite{Khachatryan:2015pea} for the simulation of fragmentation, parton shower, hadronization, and the underlying event.
In the case of the \MADGRAPH{}5\_a\MCATNLO samples, jets from the matrix element calculations are matched to the parton shower
description, following the FxFx matching prescription~\cite{Frederix:2012ps}
for the NLO samples and the MLM scheme~\cite{Mangano:2006rw} for the LO ones. The NNPDF 3.0~\cite{Ball:2014uwa} parton distribution functions (PDFs) are used for all generated samples.
Interactions of final-state particles with the CMS detector are simulated with \textsc{Geant4}~\cite{Agostinelli:2002hh}.
Simulated events include the effects of pileup,
and are weighted to reproduce the distribution of reconstructed primary vertices observed in data.

\begin{table}[!htb]
\topcaption{Monte Carlo generators used for simulating various signal and background processes.}
 \centering
 \resizebox{\textwidth}{!}{
 \begin{tabular}{|l|l|l|}
 \hline
 Process                                         & Monte Carlo generator & Perturbative   \\
                                                 &                       & order in QCD   \\
 \hline
 \Zjets                                          & \MADGRAPH{}5\_a\MCATNLO 2.2.3  & LO      \\
 \phojets                                        & \MADGRAPH{}5\_a\MCATNLO 2.2.3  & LO      \\
 \Wjets                                          & \MADGRAPH{}5\_a\MCATNLO 2.2.3  & NLO     \\
 QCD multijet                                    & \MADGRAPH{}5\_a\MCATNLO 2.2.3  & LO      \\
 \ttbar                                          & \textsc{Powheg} 2.0            & NLO     \\
 Single top quark                                & \textsc{Powheg} 2.0            & NLO     \\
 Diboson (ZZ, WZ, WW)                            & \textsc{Pythia 8.205}          & LO      \\
 \hline
 Monojet signal (spin-1 mediator)                & \textsc{Powheg} 2.0            & NLO     \\
 Monojet signal (spin-0 mediator)                & \textsc{Powheg} 2.0            & LO      \\
 Mono-V signal (spin-1 mediator)                 & \MADGRAPH{}5\_a\MCATNLO 2.2.3  & LO      \\
 Mono-V signal (scalar mediator)                 & \textsc{JHUGen} 5.2.5          & LO      \\
 \hinv~(gluon fusion)                            & \textsc{Powheg} 2.0            & NLO     \\
 \hinv~(vector boson fusion)                     & \textsc{Powheg} 2.0            & NLO     \\
 \hinv~(associated production with W or Z)       & \textsc{JHUGen} 5.2.5          & LO      \\
 \hline
 \end{tabular}
}
 \label{tab:MC_Generators}
\end{table}

\section{Event selection}\label{sec:selection}
Candidate events are selected using triggers that have thresholds of 90, 100, or 110\GeV applied equally to both \mettrig~and \mhttrig,
where \mettrig~is computed as the magnitude of the vector sum of the \pt of all the particles reconstructed at the trigger level,
and \mhttrig~is the magnitude of the vector \pt sum of jets reconstructed at the trigger level.
Jets used in the \mhttrig~computation are required to have $\pt > 20\GeV$ and $|\eta| < 5.0$.
The energy fraction attributed to neutral hadrons in these jets is required to be less than 0.9. This requirement removes jets reconstructed from detector noise.
The values of \mettrig~and \mhttrig~are calculated without including muon candidates,
allowing the same triggers to be used for selecting events in the muon control samples used for background estimation.
The trigger efficiency is measured to be about 95\% for events passing the analysis selection with $\MET \approx 200\GeV$.
The triggers become fully efficient for events with $\MET > 350\GeV$.
Events considered in this search are required to have $\MET > 200\GeV$, which ensures that the trigger efficiency is higher than 95\%.
The leading AK4 jet in the event is required to have $\pt > 100\GeV$ and $|\eta| < 2.5$. Unlike earlier searches performed by the CMS Collaboration
in this final state~\cite{Khachatryan:2014rra,Khachatryan:2016mdm}, there is no requirement on the number of reconstructed jets in the event.
The leading AK4 jet must have at least 10\% of its energy associated with charged hadrons,
and less than 80\% of its energy coming from neutral hadrons. These requirements, along with quality filters applied to tracks, muon candidates, and other objects,
reduce the background due to large misreconstructed \MET~\cite{CMS-PAS-JME-16-004}.

The dominant backgrounds in this search are the \Zvvjets~and \Wlvjets~ processes. The \Zvvjets~process constitutes the largest background and is irreducible.
The \Wlvjets~background is suppressed by vetoing events that contain at least one isolated electron or muon with $\pt > 10\GeV$, or a hadronically decaying $\tau$ lepton with $\pt > 18\GeV$.
Electron candidates must have $|\eta| < 2.5$, and are required to satisfy identification criteria based on the shower shape of the energy deposit in the ECAL, the matching of a track to
the ECAL energy cluster, and the consistency of the electron track with the primary vertex~\cite{Khachatryan:2015hwa}.
Muon candidates must have $|\eta| < 2.4$, and are required to be identified as muons by the PF algorithm.
The isolation sum of the transverse momenta of particles in a cone of radius 0.4 (0.3) around the muon (electron), corrected for the contribution of pileup,
is required to be less than 25\% (14\%) of the muon (electron) transverse momentum.
The $\tau$ lepton identification criteria~\cite{Khachatryan:2015dfa} require a jet with an identified subset of particles
whose invariant mass is consistent with that of a hadronically decaying $\tau$ lepton,
and for which the pileup-corrected isolation sum of the \pt of particle candidates within a cone of radius 0.3 around the jet axis is less than 5\GeV.
Events are vetoed if they contain an isolated photon with $\pt > 15\GeV$ that satisfies identification criteria based on its ECAL shower shape~\cite{Khachatryan:2015iwa}.
This reduces electroweak backgrounds with a photon radiated from the initial state to about 1\% of the total background.
The top quark background is suppressed by vetoing events in which
a b-jet with $\pt > 15\GeV$ is identified using the combined secondary vertex algorithm with the medium working point~\cite{Chatrchyan:2012jua,CMS-PAS-BTV-15-001},
which has a 60\% efficiency for tagging jets originating from b quarks, and a 1\% probability of misidentifying a light-flavor jet as a b-jet.
Lastly, in order to suppress the QCD multijet background in which large \MET arises from a severe mismeasurement of the jet momenta,
the minimum azimuthal angle between $\ptvecmiss$ and the directions of each of the four highest \pt AK4 jets with
$\pt > 30\GeV$ is required to be greater than 0.5 radians. The QCD multijet background is reduced to about 1\% of the total background after this requirement.

After these criteria are applied, events are classified into mono-V or monojet categories.
If a V boson has $\pt > 250\GeV$, its hadronic decay is more likely to be reconstructed as a single AK8 jet than as two AK4 jets.
An event is categorized as a mono-V event if it has $\MET > 250\GeV$, and the leading AK8 jet in the event
has $\pt > 250\GeV$ and $|\eta| < 2.4$, and also passes requirements used to identify jets arising from hadronic decays of Lorentz-boosted V bosons.
Jets arising from hadronic decays of a V boson are identified using the $N$-subjettiness variable $\tau_N$~\cite{Thaler:2010tr}. Low values of $\tau_N$ are indicative of an $N$-prong decay. In particular, the ratio $\tau_2 / \tau_1$ discriminates the two-prong decays of a V boson from QCD jets, and the leading AK8 jet is required to have $\tau_2 / \tau_1 < 0.6$. Additionally,
the invariant mass of the jet is required to be between 65 and 105\GeV in order to be consistent with the mass of the W or Z boson.
The jet mass is computed after pruning~\cite{Ellis:2009me}, which involves reclustering of the jet constituents using the Cambridge--Aachen algorithm~\cite{Dokshitzer:1997in,Wobisch:1998wt}
and removing the soft constituents in every recombination step, thereby improving the jet mass resolution.  The requirements on the $\tau_2 / \tau_1$ ratio and the jet mass result in a 70\%
efficiency for tagging jets originating from V bosons, and a 5\% probability of misidentifying a QCD jet as a V jet. If an event fails any of these mono-V selection requirements,
it is assigned to the monojet category. The selection requirements for the mono-V and monojet categories are listed in Table~\ref{tab:Event_Selection}.

\begin{table}[!htb]
\topcaption{Selection requirements for the mono-V and monojet event categories.}
 \centering
 \resizebox{\textwidth}{!}{
 \begin{tabular}{|l|c|c|}
 \hline
 Variable                                                   & Mono-V                           & Monojet   \\
                                                            & requirement                      & requirement   \\
 \hline
 \MET                                                       & $ > 250$ \GeV                    & $ > 200$ \GeV         \\
 Leading AK4 jet $\pt$                                      & \multicolumn{2}{c|}{$ > 100$ \GeV}                       \\
 Leading AK4 jet $|\eta|$                                   & \multicolumn{2}{c|}{$ < 2.5$}                            \\
 Charged hadron energy fraction of leading AK4 jet          & \multicolumn{2}{c|}{$ > 0.1$}                            \\
 Neutral hadron energy fraction of leading AK4 jet          & \multicolumn{2}{c|}{$ < 0.8$}                            \\
 Number of muons ($\pt > 10 ~\GeV, |\eta| < 2.4$)           & \multicolumn{2}{c|}{$0$}                                 \\
 Number of electrons ($\pt > 10 ~\GeV, |\eta| < 2.5$)       & \multicolumn{2}{c|}{$0$}                                 \\
 Number of $\tau$ leptons ($\pt > 18 ~\GeV, |\eta| < 2.3$)  & \multicolumn{2}{c|}{$0$}                                 \\
 Number of photons ($\pt > 15 ~\GeV, |\eta| < 2.5$)         & \multicolumn{2}{c|}{$0$}                                 \\
 Number of b jets ($\pt > 15 ~\GeV, |\eta| < 2.4$)          & \multicolumn{2}{c|}{$0$}                                 \\
 $\Delta\phi$ between four highest \pt jets and \MET        & \multicolumn{2}{c|}{$ > 0.5$ radians}                    \\
 \hline
 Leading AK8 jet $\pt$                                      & $ > 250 $ \GeV                   &                       \\
 Leading AK8 jet $\eta$                                     & $ < 2.4 $                        & Fails any of the mono-V  \\
 Leading AK8 jet $\tau_2 / \tau_1$                          & $ < 0.6 $                        & AK8 jet requirements  \\
 Leading AK8 jet mass ($m_{\mathrm{J}}$)                    & $65 < m_{\mathrm{J}} < 105$ \GeV &                       \\
 \hline
 \end{tabular}
}
 \label{tab:Event_Selection}
\end{table}

\section{Background estimation}
The \Zvvjets~and \Wlvjets~processes constitute about 90\% of the total background in this search.
These background contributions are estimated using data from dimuon, dielectron, single-muon, single-electron, and \phojets~control samples.
Events in each of these control samples are further classified into the monojet and mono-V categories, resulting in ten mutually exclusive control samples.
The \MET in the control samples is redefined by excluding the leptons and the photons from the calculation.
The \pt of the resulting hadronic recoil system resembles the \MET distribution of the electroweak backgrounds in the signal region.
Therefore, the hadronic recoil \pt is used as a proxy for \MET in the control regions.

The dimuon and single-muon events are selected with the same \MET triggers that are used to select the signal events.
The dimuon events are required to contain exactly two oppositely charged muons, each with $\pt > 10\GeV$.
Events are vetoed if there is an additional muon or electron with $\pt > 10\GeV$.
At least one of the two muons is required to have $\pt > 20\GeV$ and to pass tight identification requirements
based on the number of measurements in the tracker and the muon system, the quality of the muon track fit, and the
consistency of the muon track with the primary vertex. The isolation sum of the \pt of particles in a cone of radius
0.4 around the muon, corrected for the contribution of pileup, is required to be less than 15\% of the muon \pt.
The invariant mass of the dimuon system is required to be between 60 and 120\GeV, in order to be consistent with a Z boson decay.
The single-muon events are required to contain exactly one tightly identified and isolated muon with $\pt > 20\GeV$.
No additional muon or electron with $\pt > 10\GeV$ is allowed, and the transverse mass of the muon-\MET system is required to be less
than 160\GeV. The transverse mass (\mt) is computed as $\mt^{2} = 2\MET \pt^{\mu} (1 - \mathrm{cos}\Delta\phi)$, where $\pt^{\mu}$
is the \pt of the muon, and $\Delta\phi$ is the angle between $\pt^{\mu}$ and $\ptvecmiss$.
The dimuon and single-muon events are further required to satisfy all other selection requirements imposed on the
signal events with the \MET replaced by the \pt of the hadronic recoil system. The distribution of the hadronic recoil \pt is then used to estimate the \Zvvjets~and \Wlvjets~backgrounds in the signal region.

The dielectron control sample is constructed using events with exactly two oppositely charged electrons with $\pt > 10\GeV$, and no additional muon or electron.
The invariant mass of the dielectron system is required to be between 60 and 120\GeV, as in the case of the dimuon events.
A single-electron trigger with a \pt threshold of 27\GeV is used to select these events.
If the Z boson has $\pt > 600\GeV$, the two electrons produced in its decay typically have a small angular separation,
and are likely to be included in each other's isolation cones. This effect results in some inefficiency for the single-electron trigger, which imposes isolation requirements
on electron candidates. In order to overcome this inefficiency, events are also accepted if they pass a single-electron trigger that has a \pt threshold of 105\GeV
and no isolation requirements on the electron candidate. Furthermore, in order to improve the efficiency of the electron triggers in the early part of the data taking,
additional events passing a trigger with a threshold of 800\GeV on the total sum of the \pt of jets (\HT) reconstructed at the trigger level are also included.
The same set of triggers is also used for selecting events in the single-electron control sample.
At least one of the two electrons in the dielectron control sample is required to have $\pt > 40\GeV$, and is required to pass tight identification requirements on the
shower shape of its ECAL energy deposit, the matching of a track to the ECAL energy cluster, and the consistency of the electron track with the primary vertex.
The isolation sum of the \pt of particles in a cone of radius 0.3 around this electron, corrected for the contribution of pileup,
is required to be less than 3.5\% of the electron \pt for electrons within the ECAL barrel ($|\eta| < 1.48$), and less than 6.5\% of the electron \pt for
electrons within the ECAL endcaps ($1.48 < |\eta| < 2.50$). The single-electron events are required to contain exactly one tightly identified and isolated electron with $\pt > 40\GeV$.
No additional muons or electrons with $\pt > 10\GeV$ are allowed. The QCD background in the single-electron control sample is suppressed by requiring \MET$ > 50\GeV$,
and $\mt < 160 \GeV$.

The \phojets~control sample is constructed using events with one high-\pt photon that are selected using single-photon triggers with \pt thresholds of 165 or 175\GeV.
As in the case of the electron control samples, additional events passing the \HT trigger with a threshold of 800\GeV are also included.
The photon \pt is required to be larger than 175\GeV, which ensures that the trigger efficiency is greater than 98\%. The photon candidate
is required to be reconstructed in the ECAL barrel, and is required to pass identification and isolation criteria that ensure an efficiency of 80\%
in selecting prompt photons, and a sample purity of 95\% \cite{Khachatryan:2015iwa}.

The procedure for estimating the \Zvvjets~and \Wlvjets~backgrounds relies on transfer factors derived from simulation that connect the yields
of electroweak processes in the control samples with the background estimates in the signal region, for a given range of \MET.
The transfer factors for the dilepton control samples relate the yields of \Zmm~and \Zee~events to the \Zvv~background in the signal region
by taking into account the difference in the branching fractions of \Zvv~and \Zll~decays and the effect of lepton acceptance and selection efficiencies.
In the case of dielectron events these transfer factors also account for the difference in efficiencies of the electron and \MET triggers.
The transfer factor for the \phojets~control sample takes into account the difference in the cross sections of the \phojets~and \Zvvjets~processes, the effect of photon acceptance and efficiency,
and the difference in the efficiencies of the photon and \MET triggers. Transfer factors are also defined between the \Wmn~and \Wen~event yields in the single-lepton control samples and
the \Wlvjets~background estimate in the signal region. These take into account the effect of lepton acceptance, lepton selection efficiencies, $\tau$ lepton veto efficiency, and the
difference in trigger efficiencies in the case of the single-electron control sample.
Finally, a transfer factor is also defined to connect the \Zvvjets~and \Wlvjets~background yields in the signal region.
The photon transfer factor relies on an accurate estimate of the ratio of the \phojets~and \Zjets~cross sections. Similarly, the transfer factor
between the \Zvvjets~and \Wlvjets~backgrounds relies on an accurate prediction of the ratio of the \Wjets~and \Zjets~cross sections.
Therefore, the LO simulations for the \Zjets~and \phojets~processes are corrected using \pt-dependent NLO QCD K-factors derived
using \MADGRAPH{}5\_a\MCATNLO, and the \Zjets, \Wjets, and \phojets~processes are corrected using \pt-dependent NLO electroweak K-factors
from theoretical calculations~\cite{Kuhn:2005gv,Kallweit:2014xda,Kallweit:2015dum}.

The \Zvvjets~and \Wlvjets~background yields are determined through a maximum likelihood fit, performed simultaneously across all the bins of hadronic recoil \pt
in the ten control samples and \MET in the two signal regions. The likelihood function $\mathcal{L}_k$ for each of the two event categories $k$, corresponding to the monojet and mono-V selections,
is defined as
\begin{equation}
\begin{aligned}
\mathcal{L}_k(\boldsymbol{\mu}^{\Zvv}, \boldsymbol{\mu}, \boldsymbol{\theta}) =&
\prod_{i} \mathrm{Poisson}\left(d^{\gamma}_{i} |B^{\gamma}_{i}(\boldsymbol{\theta}) +\frac{ \muz_{i} }{R^{\gamma}_{i}(\boldsymbol{\theta})}   \right) \\
&\times \prod_{i} \mathrm{Poisson}\left(d^{\mu\mu}_{i}|B^{\mu\mu}_{i}(\boldsymbol{\theta}) +\frac{\muz_{i} }{R^{\mu\mu}_{i}     (\boldsymbol{\theta})}   \right ) \\
&\times \prod_{i} \mathrm{Poisson}\left(d^{\mathrm{ee}}_{i}|B^{\mathrm{ee}}_{i}(\boldsymbol{\theta}) +\frac{\muz_{i} }{R^{\mathrm{ee}}_{i}     (\boldsymbol{\theta})}   \right ) \\
&\times \prod_{i} \mathrm{Poisson}\left(d^{\mu}_{i}|B^{\mu}_{i}(\boldsymbol{\theta}) +\frac{f_{i}(\boldsymbol{\theta})\muz_{i}}{R^{\mu}_{i}(\boldsymbol{\theta})} \right)\\
&\times \prod_{i} \mathrm{Poisson}\left(d^{\mathrm{e}}_{i}|B^{\mathrm{e}}_{i}(\boldsymbol{\theta}) +\frac{f_{i}(\boldsymbol{\theta})\muz_{i}}{R^{\mathrm{e}}_{i}(\boldsymbol{\theta})} \right)\\
&\times \prod_{i} \mathrm{Poisson}\left(d_{i}     |B_{i}(\boldsymbol{\theta}) + (1+f_{i}(\boldsymbol{\theta})) \muz_{i}  + \boldsymbol{\mu} S_{i}(\boldsymbol{\theta})\right )  \label{eqn:candclh} \\
\end{aligned}
\end{equation}
where $\mathrm{Poisson}(x|y) = y^{x}\re^{-y}/x!$.
The symbols $d^{\gamma}_{i},d^{\mu\mu}_{i},d^{\mathrm{ee}}_{i},d^{\mu}_{i},d^{\mathrm{e}}_{i}$, and $d_{i}$ denote the observed number of events in
each bin $i$ of the \phojets, dimuon, dielectron, single-muon, and single-electron control samples, and the signal region, respectively.
The symbol $f_{i}$ denotes the transfer factor between the \Zvvjets~and \Wlvjets~backgrounds in the signal region, and represents a constraint between these backgrounds.
The symbols $R^{\gamma}_{i},R^{\mu\mu}_{i},R^{\mathrm{ee}}_{i},R^{\mu}_{i}$, and $R^{\mathrm{e}}_{i}$ are the transfer factors from the
\phojets, dimuon, dielectron, single-muon, and single-electron control samples, respectively, to the signal region; the contributions from other background processes
in these control samples are denoted by $B^{\gamma}_{i},B^{\mu\mu}_{i},B^{\mathrm{ee}}_{i},B^{\mu}_{i}$, and $B^{\mathrm{e}}_{i}$,
respectively. The parameter $\muz_i$ represents the yield of the \Zvvjets~background in each bin $i$ of \MET in the signal region, and this parameter is left floating in the fit.
The likelihood also includes a term for the signal region in which $B_{i}$ represents all the backgrounds apart from \Zvvjets~and \Wlvjets,
$S_i$ represents the nominal signal prediction, and $\boldsymbol{\mu}$ denotes the signal strength parameter.
The systematic uncertainties are modeled as nuisance parameters ($\boldsymbol{\theta}$).

\begin{figure*}[hbtp]\centering
\includegraphics[width=0.46\textwidth]{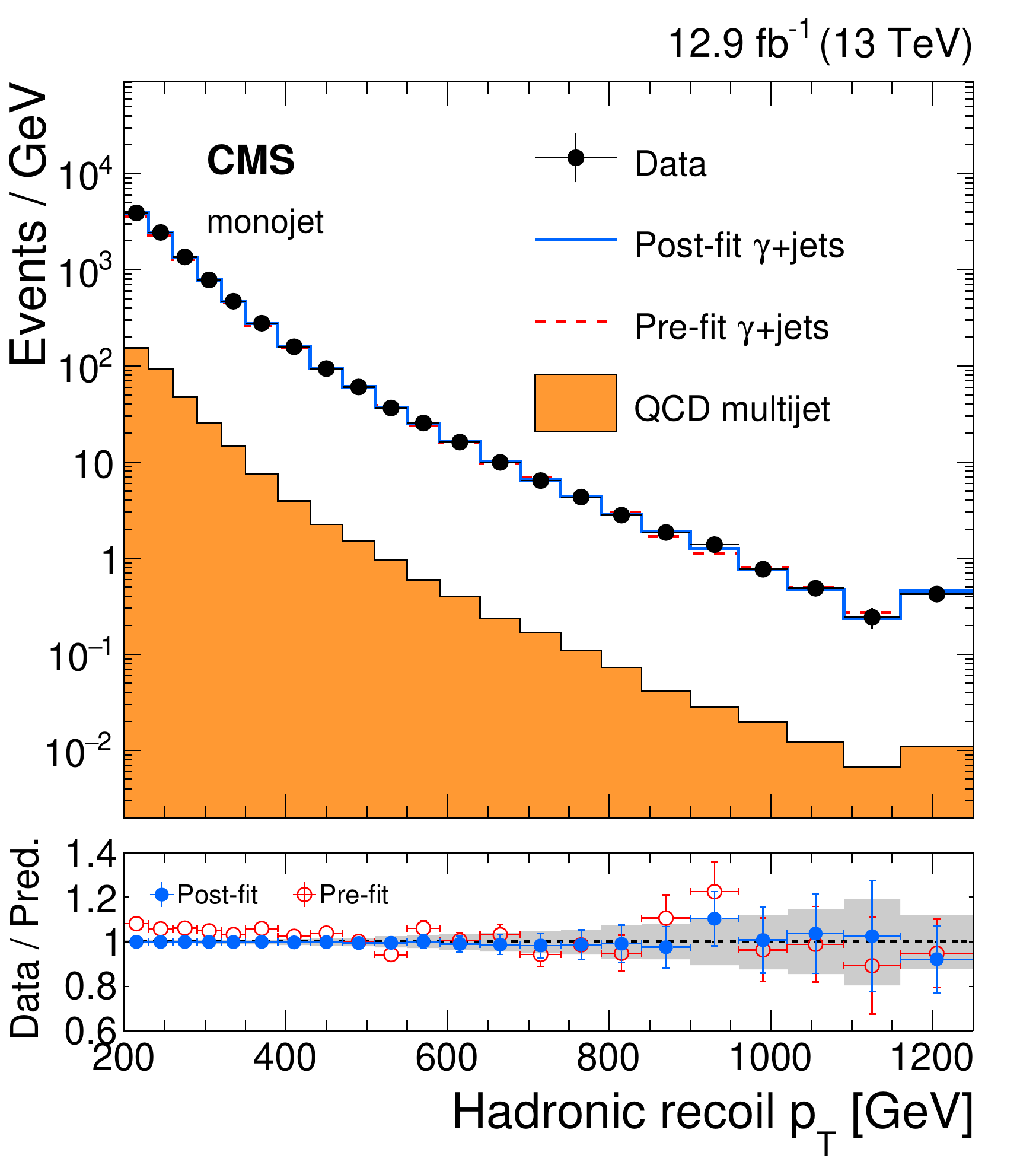}
\includegraphics[width=0.46\textwidth]{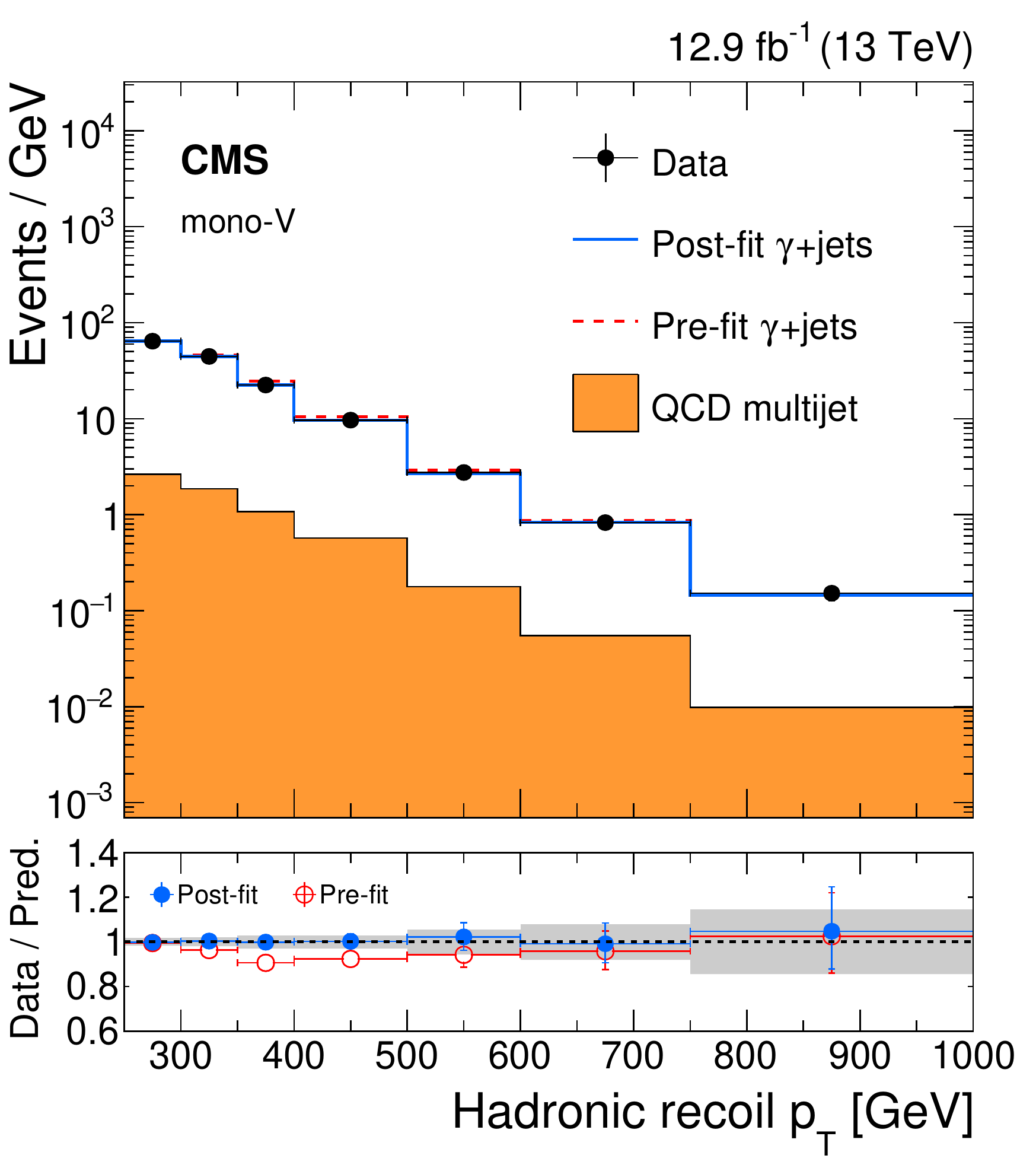}
\caption{
Comparison between data and Monte Carlo simulation in the \phojets~control sample before and after performing the simultaneous fit across all the control samples and the signal region,
assuming the absence of any signal. The left plot shows the monojet category and the right plot shows the mono-V category.
The hadronic recoil \pt in \phojets~events is used as a proxy for \MET in the signal region.
The filled histogram indicates the multijet background. Ratios of data and the pre-fit background prediction (red points) and post-fit background prediction (blue points)
are shown for both the monojet and mono-V signal categories. The gray band indicates the overall post-fit uncertainty.
The last bin includes all events with hadronic recoil \pt larger than 1160 (750)\GeV in the monojet (mono-V) category.
}
\label{fig:gamCR}\end{figure*}

\begin{figure*}[hbtp]\centering
\includegraphics[width=0.46\textwidth]{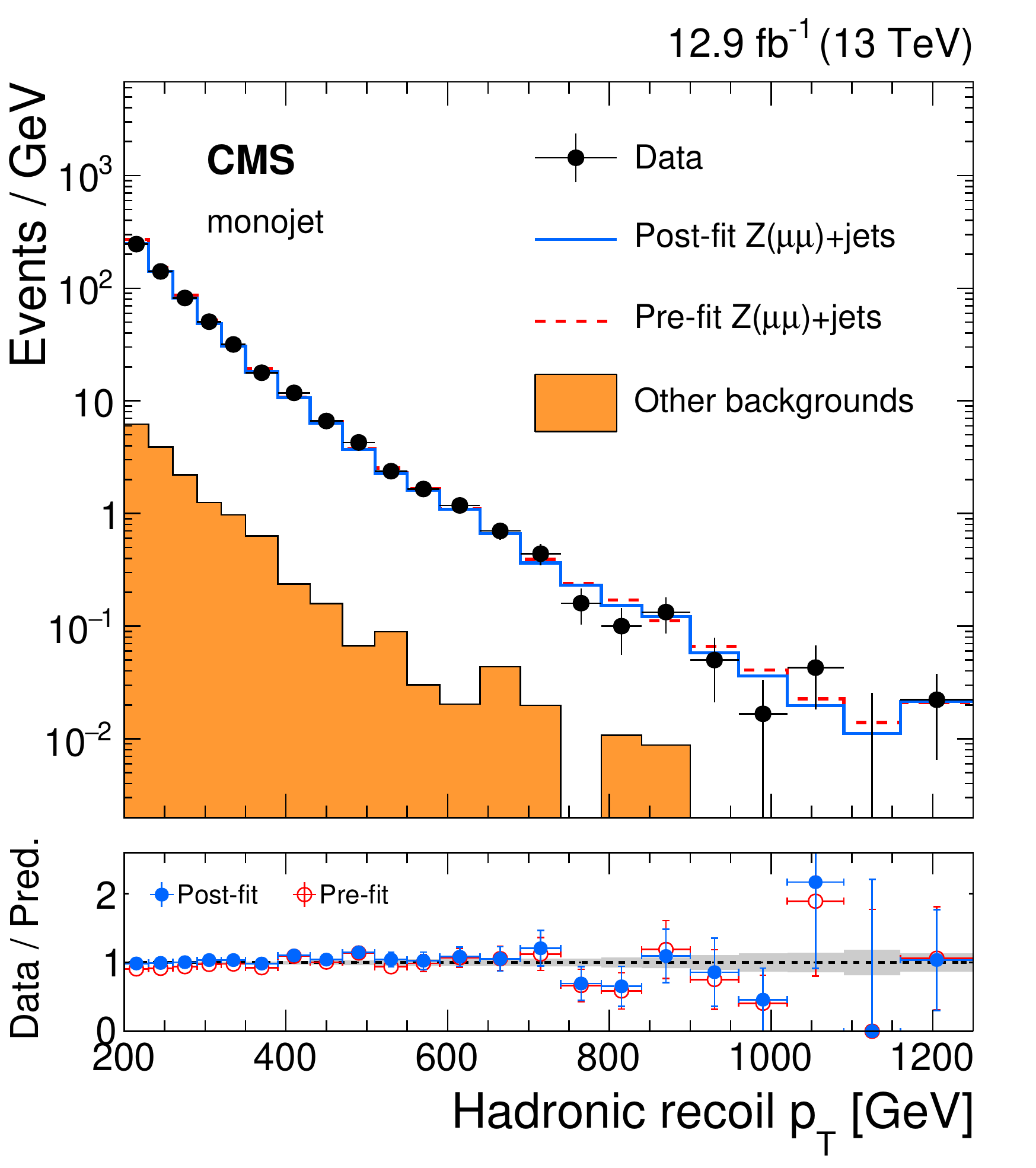}
\includegraphics[width=0.46\textwidth]{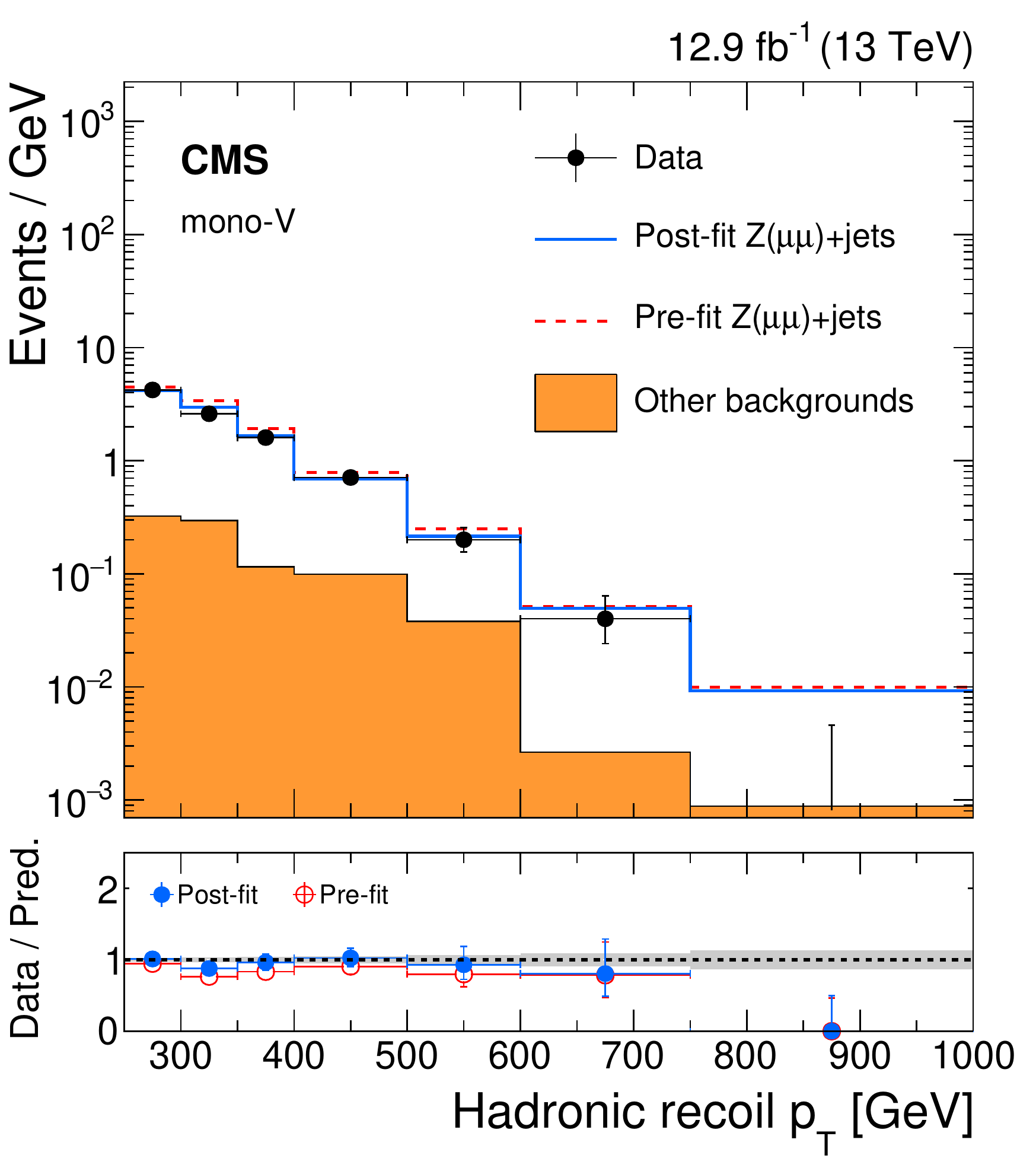}
\includegraphics[width=0.46\textwidth]{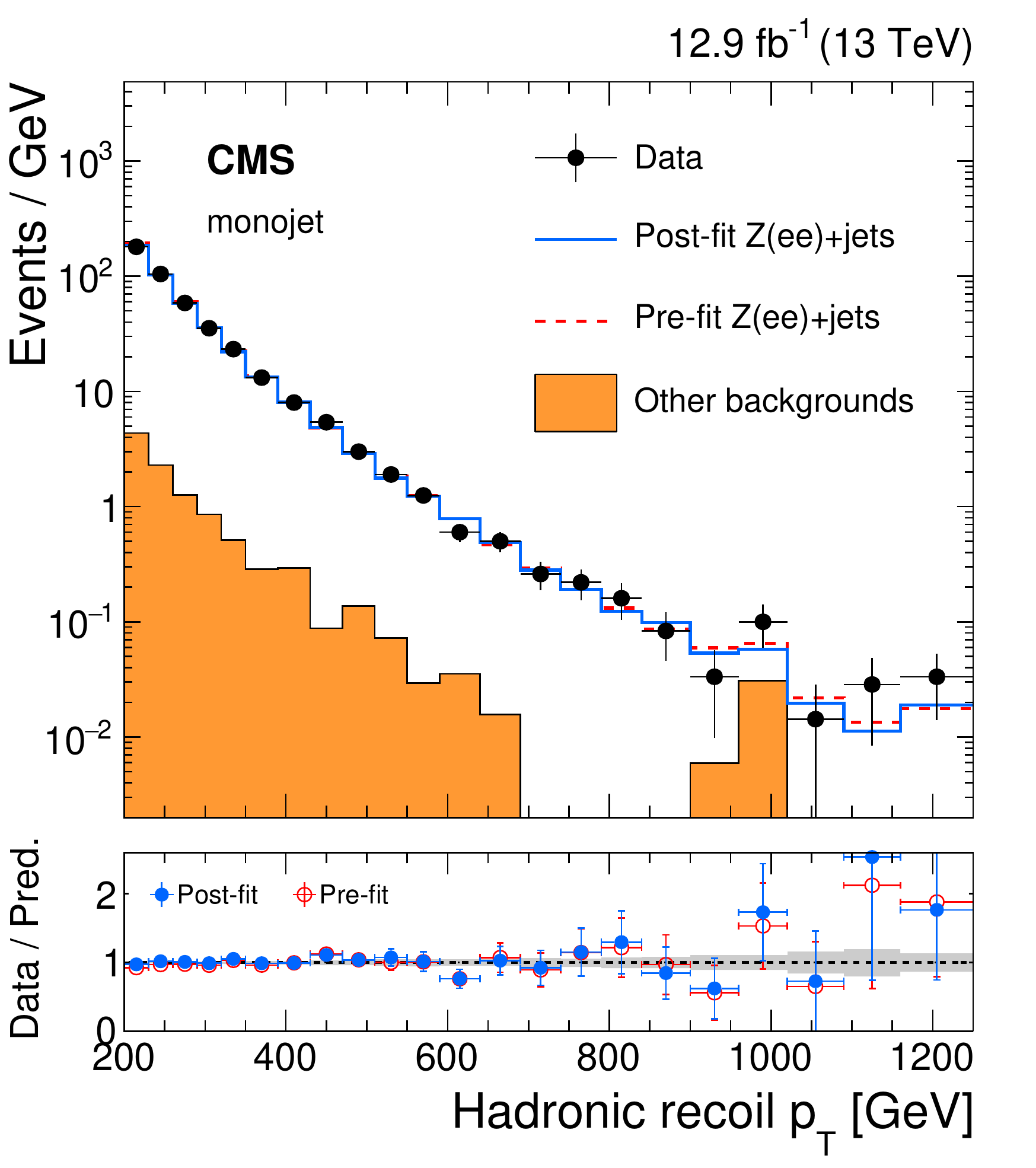}
\includegraphics[width=0.46\textwidth]{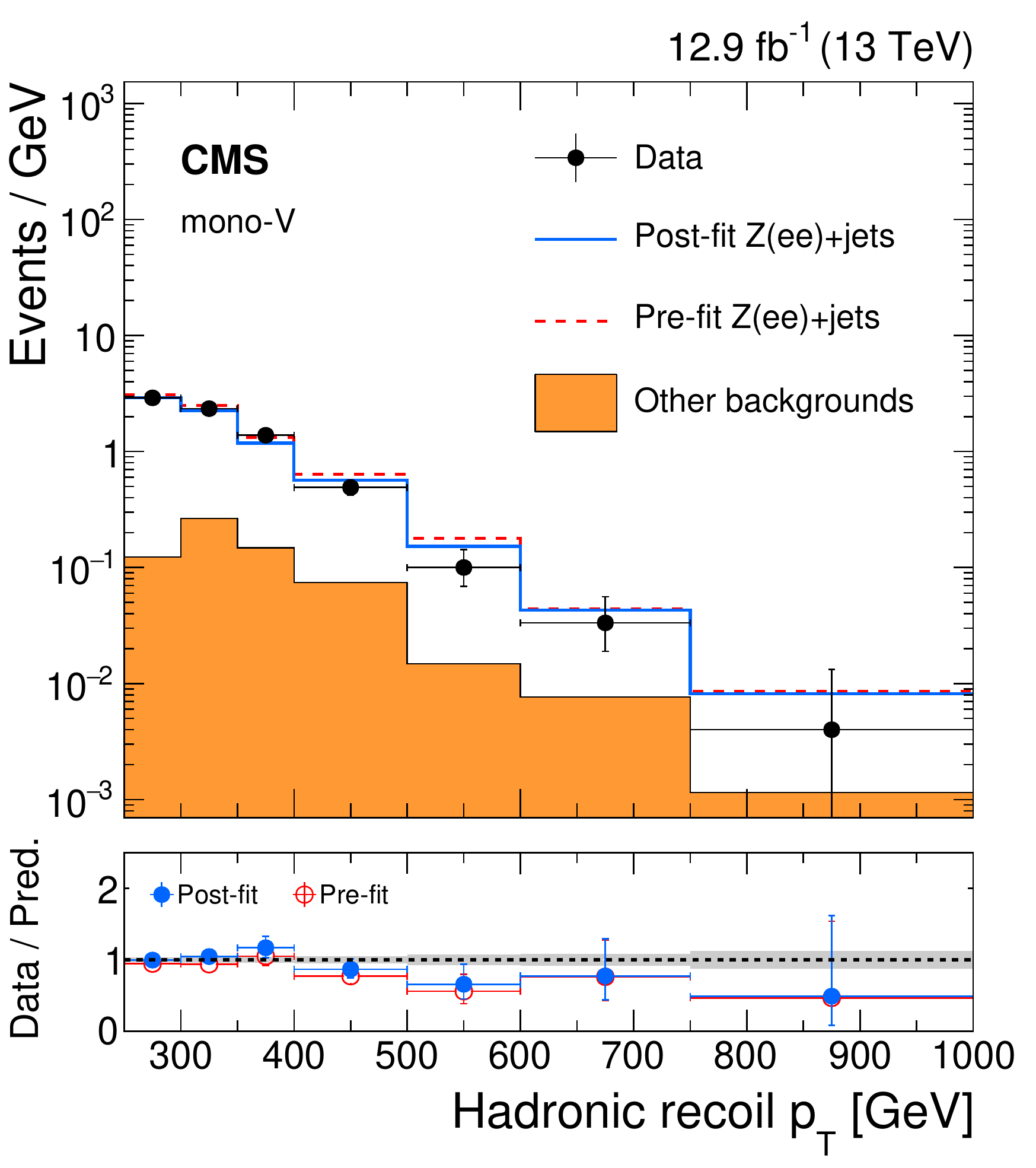}
\caption{
Comparison between data and Monte Carlo simulation in the dilepton control samples before and after performing the simultaneous fit across all the control samples and the signal region,
assuming the absence of any signal. Plots on the upper left and right correspond to the monojet and mono-V categories, respectively, in the dimuon control sample.
Plots on the bottom left and right correspond to the monojet and mono-V categories, respectively, in the dielectron control sample.
The hadronic recoil \pt in dilepton events is used as a proxy for \MET in the signal region.
The filled histogram indicates all processes other than \Zlljets. Ratios of data and the pre-fit background prediction (red points) and post-fit background prediction (blue points)
are shown for both the monojet and mono-V signal categories. The gray band indicates the overall post-fit uncertainty.
The last bin includes all events with hadronic recoil \pt larger than 1160 (750)\GeV in the monojet (mono-V) category.
}
\label{fig:zllCR}\end{figure*}

\begin{figure*}[hbtp]\centering
\includegraphics[width=0.46\textwidth]{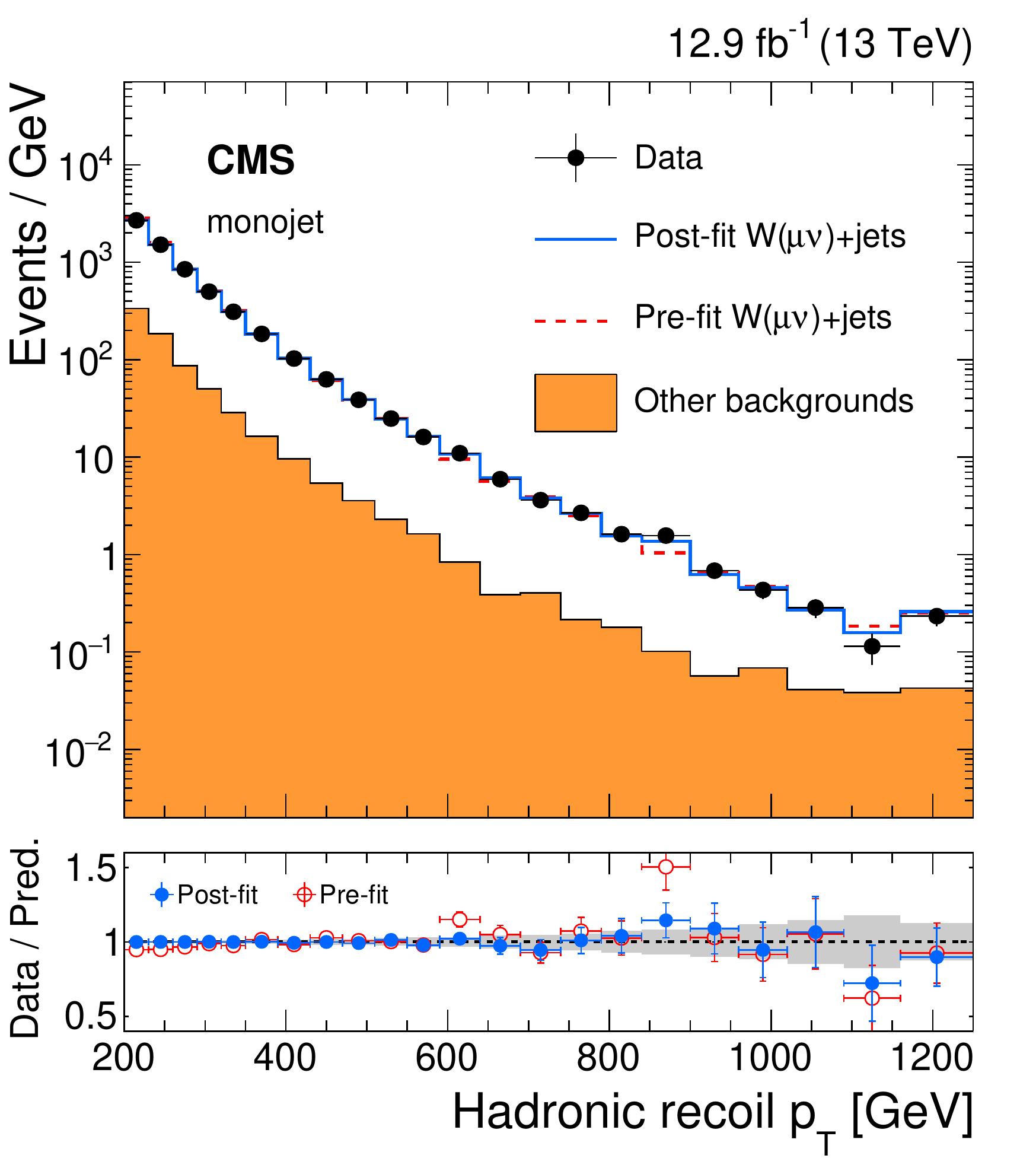}
\includegraphics[width=0.46\textwidth]{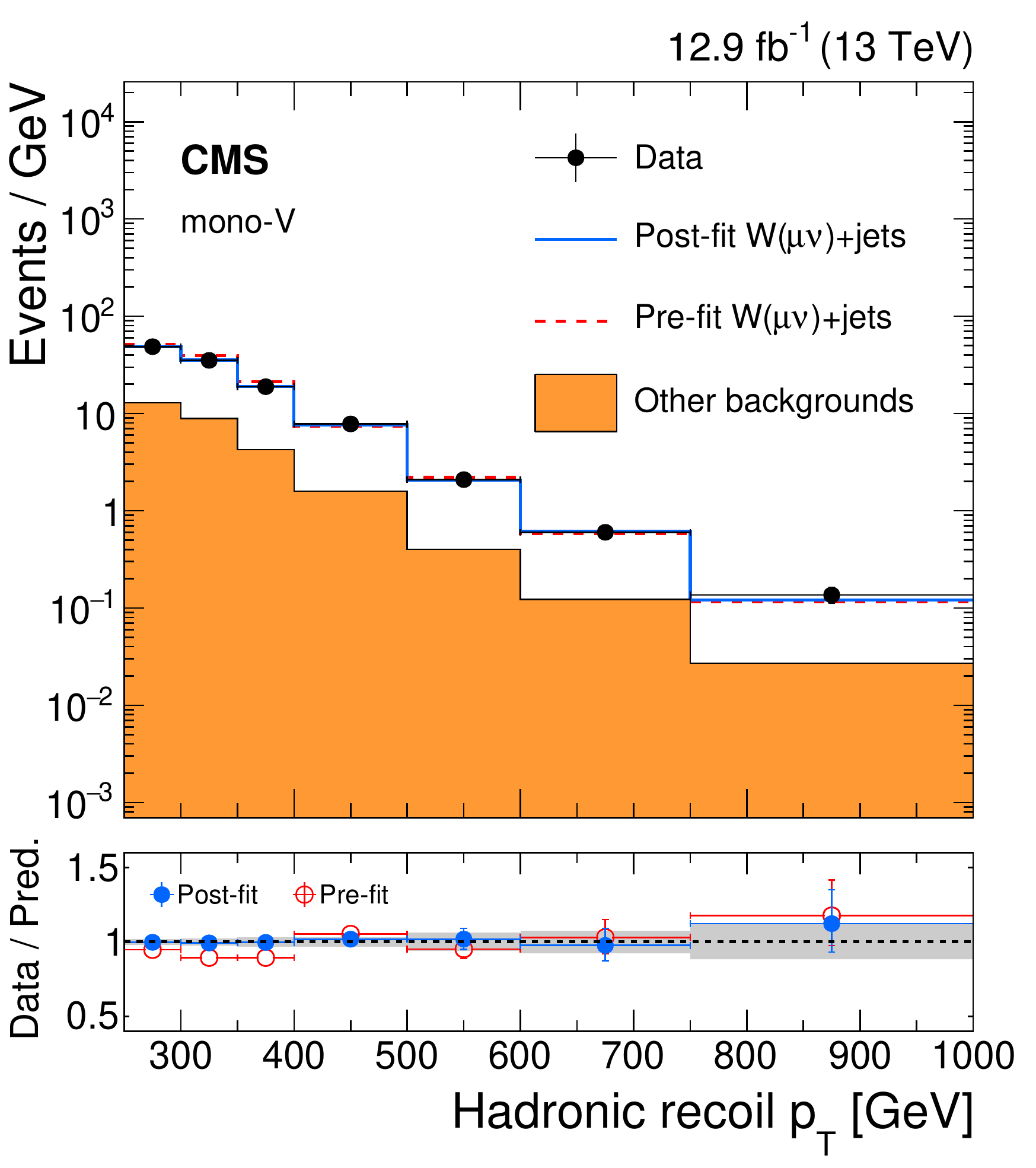}
\includegraphics[width=0.46\textwidth]{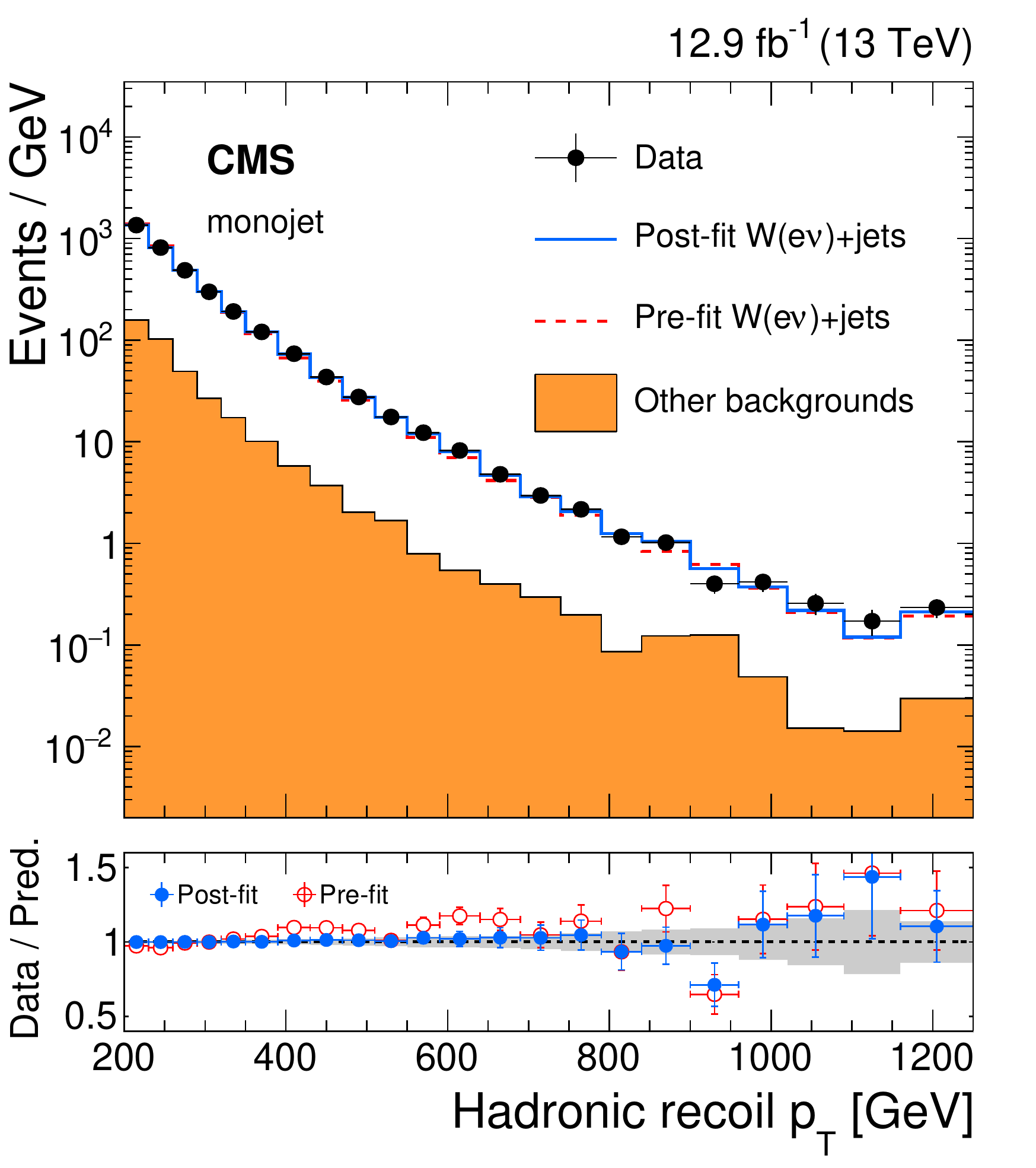}
\includegraphics[width=0.46\textwidth]{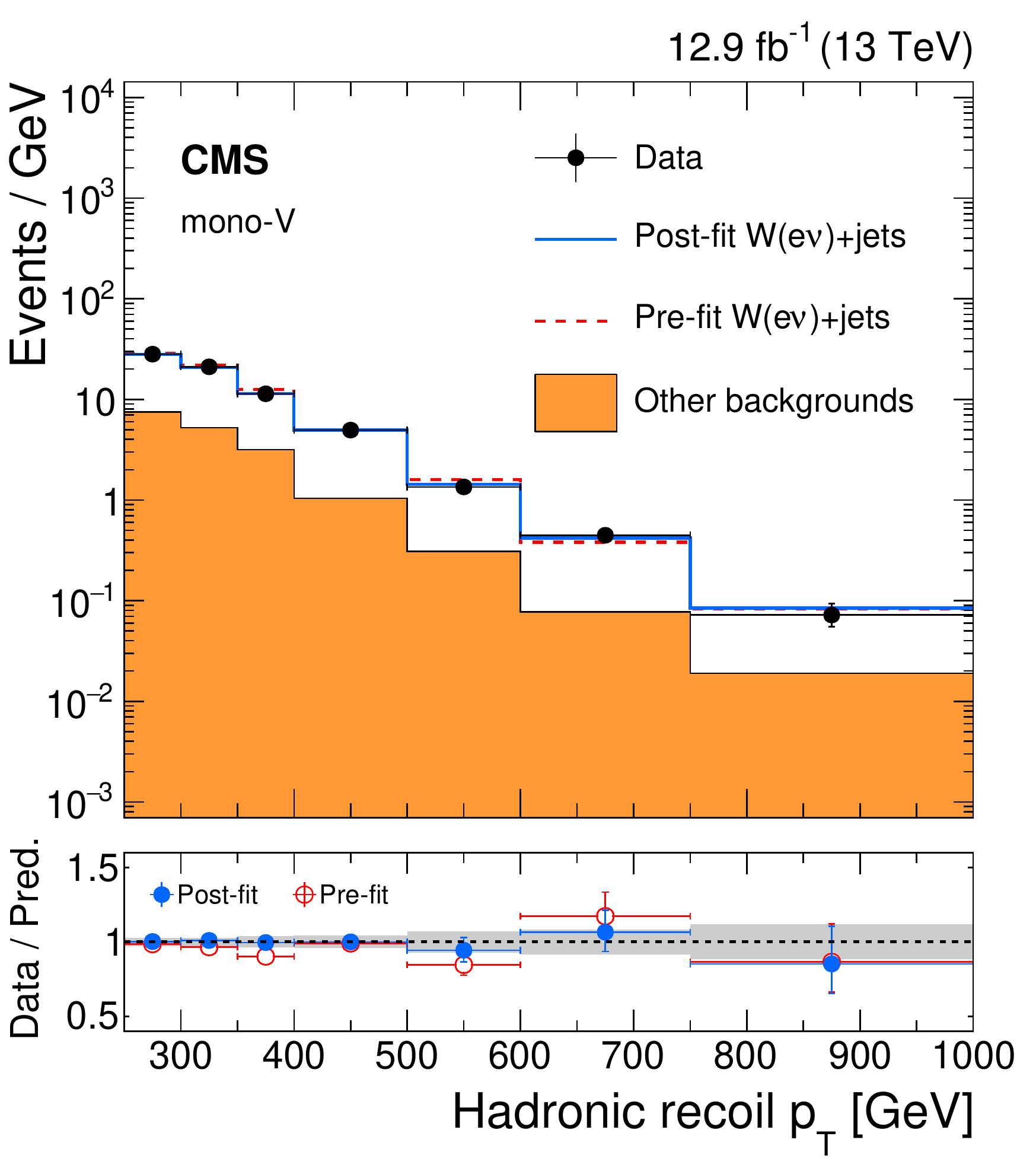}
\caption{
Comparison between data and Monte Carlo simulation in the single-lepton control samples before and after performing the simultaneous fit across all the control samples and
the signal region, assuming the absence of any signal. Plots on the upper left and right correspond to the monojet and mono-V categories, respectively, in the single-muon control sample.
Plots on the bottom left and right correspond to the monojet and mono-V categories, respectively, in the single-electron control sample.
The hadronic recoil \pt in single-lepton events is used as a proxy for \MET in the signal region.
The filled histogram indicates all processes other than \Wlvjets. Ratios of data and the pre-fit background prediction (red points) and post-fit background prediction (blue points)
are shown for both the monojet and mono-V signal categories. The gray band indicates the overall post-fit uncertainty.
The last bin includes all events with hadronic recoil \pt larger than 1160 (750)\GeV in the monojet (mono-V) category.
}
\label{fig:wlnCR}\end{figure*}

The uncertainties in the \Zvvjets~and \Wlvjets~backgrounds enter the likelihood as constrained
perturbations of the transfer factors $R^{\gamma}_{i},R^{\mu\mu}_{i},R^{\mathrm{ee}}_{i},R^{\mu}_{i},R^{\mathrm{e}}_{i}$ and $f_i$.
These include theoretical uncertainties in the \phojets~to \Zjets, and \Wjets~to \Zjets~differential cross section ratios
from the choice of the renormalization (10--15\%) and factorization
(1--10\%) scales~\cite{Khachatryan:2016mdm}, and the PDF modeling uncertainty, which is found to be negligible. The effect of missing higher-order electroweak corrections to the
\phojets, \Wjets, and \Zjets~processes is covered by propagating the full NLO electroweak correction as a function of the boson \pt as the uncertainty.
The resulting uncertainty varies within 2--14\% and 1--9\% for the \phojets~to \Zjets~and \Wjets~to \Zjets~differential cross section ratios,
respectively, and it is conservatively considered to be uncorrelated across the bins of hadronic recoil \pt. Uncertainties in the reconstruction efficiencies
of leptons (1\% per muon or electron); in selection efficiencies of leptons (2\% per muon or electron), photons (2\%), and hadronically decaying $\tau$ leptons (3\%);
in the purity of photons in the~\phojets~control sample (2\%); and in the efficiency of the electron (2\%), photon (2\%), and \MET (1\%) triggers, are included and their correlations
across all the bins of hadronic recoil \pt are taken into account.
Figures~\ref{fig:gamCR}--\ref{fig:wlnCR} show the results of the combined fit in the ten control samples and the two signal regions assuming the absence of any signal.
Data in the control samples are compared to the pre-fit predictions from simulation and the post-fit estimates obtained after performing the fit.
The control samples with larger yields dominate the fit results.

In addition to the \Zvvjets~and \Wlvjets~processes, several other sources of background contribute to the total event yield in the signal region.
These include QCD multijet events that have little genuine \MET. However, jet mismeasurement and instrumental
effects can give rise to high \MET tails. A $\Delta\phi$ extrapolation method~\cite{Collaboration:2011ida} is used to estimate this background.
In this method, a background-enriched control sample is obtained by selecting events that fail the $\Delta\phi$ requirement between jets and \MET, but pass the  remaining
event selection criteria. An estimate of the multijet background in the signal region is obtained
by applying \MET-dependent transfer factors, derived from simulated QCD multijet events, to this control sample.
The overall uncertainty in the multijet background estimate, based on the variations of the jet response and the statistical uncertainties in the transfer factors,
ranges from 50 to 150\%, depending on the event category and the \MET region.

The remaining background sources include top quark and diboson processes, which are estimated directly from simulation.
The \pt distribution of the top quark in simulation is corrected to match the observed \pt distribution in data~\cite{Czakon:2015owf}.
A systematic uncertainty of 10\% is assigned to the prediction of the top quark background cross section~\cite{Khachatryan:2015uqb}.
An additional 10\% uncertainty is assigned to the top quark background normalization to take account of the
modeling of the top quark \pt distribution in simulation. The overall normalization of the diboson background has an uncertainty of 20\%~\cite{Khachatryan:2016txa,Khachatryan:2016tgp}.
These uncertainties in the top quark and diboson backgrounds are correlated across the signal and control samples.
Several experimental sources of uncertainty are associated with the backgrounds estimated from simulation.
An uncertainty of 6.2\% in the integrated luminosity measurement~\cite{CMS-PAS-LUM-15-001} is propagated to the background yields.
The uncertainty in the efficiency of the b-jet veto is estimated to be 6\% for the top quark background and 2\% for the diboson background.
The uncertainty in the efficiency of the V tagging requirements is estimated to be 13\% in the mono-V category.
The uncertainty in the modeling of \MET in simulation~\cite{Khachatryan:2014gga} is dominated by the jet energy scale uncertainty, and is estimated to be 5\%.

\section{Results and interpretation}
\begin{figure*}[!h]
\centering
\includegraphics[width=0.46\textwidth]{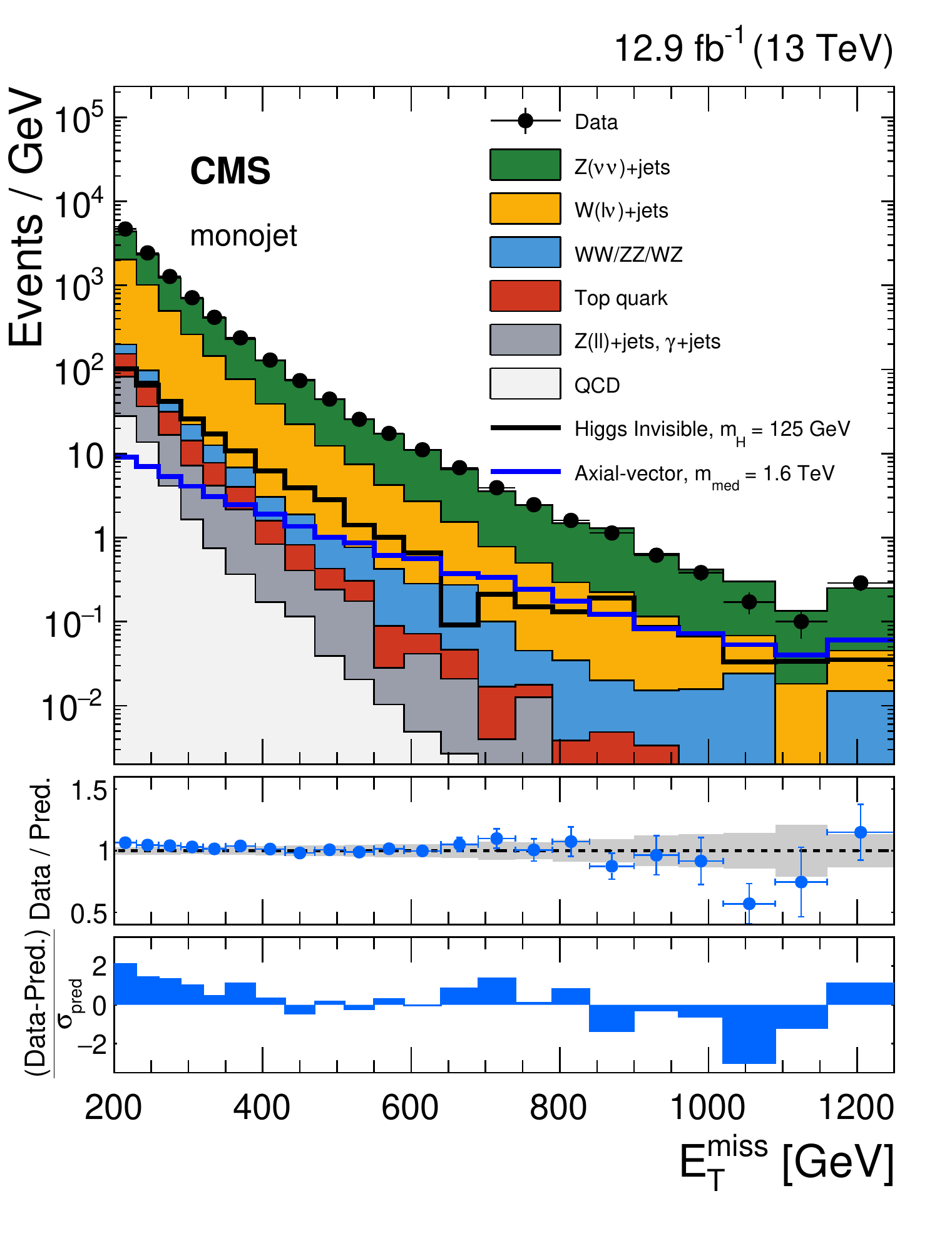}
\includegraphics[width=0.46\textwidth]{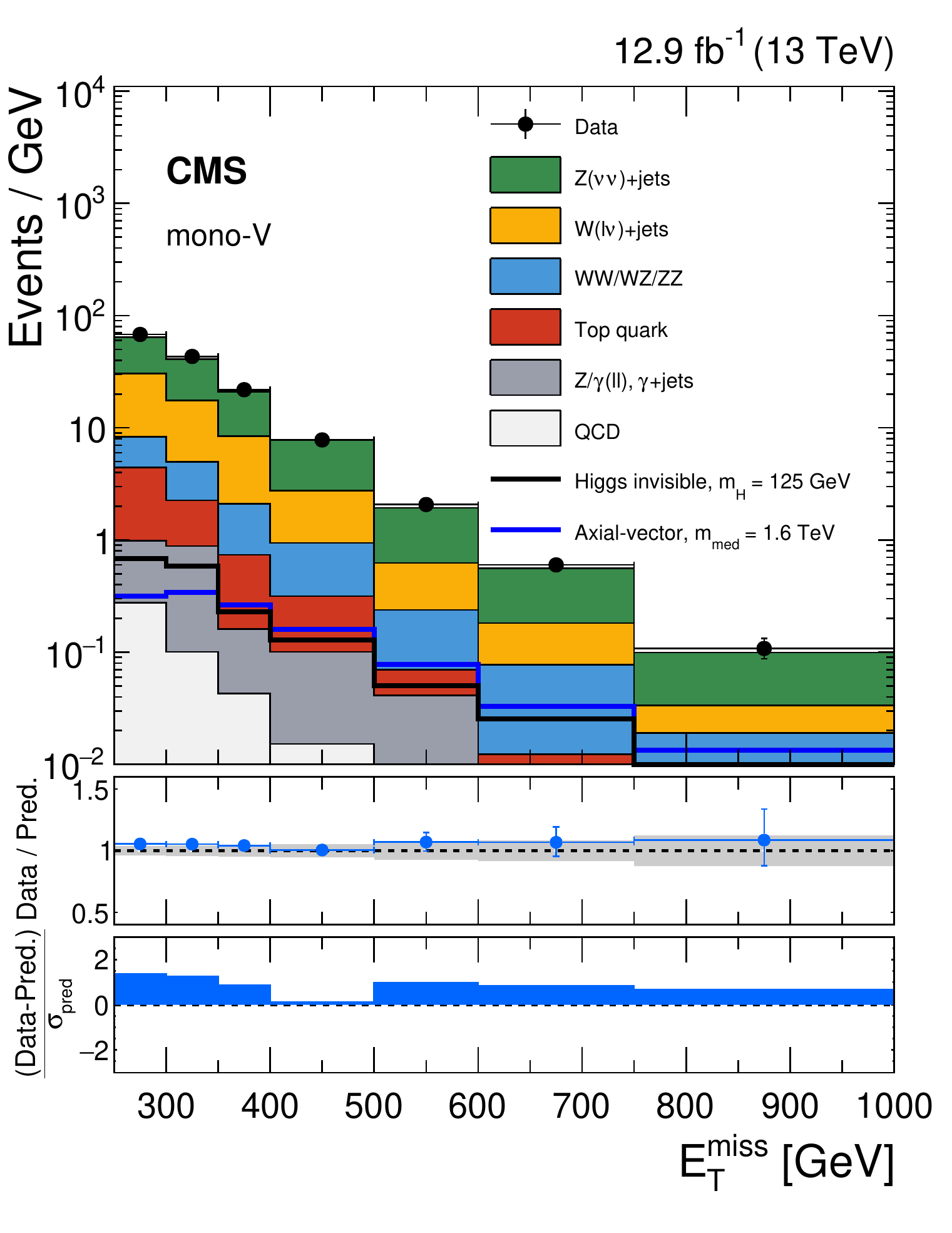}
\caption{
Observed \MET distribution in the monojet (left) and mono-V (right) signal regions compared with the background expectations for various SM processes
evaluated after performing a combined fit to the data in all the control samples, but excluding the signal region.
The last bin includes all events with $\MET > $ 1160 (750)\GeV for the monojet (mono-V) category.
Expected signal distributions for a 125\GeV Higgs boson decaying exclusively to invisible particles, and for a 1.6\TeV axial-vector mediator decaying to 1\GeV DM particles, are overlaid.
The ratio of data and the post-fit background prediction is shown for both the monojet and mono-V signal regions.
The gray bands in these ratio plots indicate the post-fit uncertainty in the background prediction.
Finally, the distributions of the pulls, defined as the difference between data and the post-fit background prediction relative to the post-fit uncertainty in the prediction,
are also shown in the lower panels.
}
\label{fig:moneyplots_SRmask}

\end{figure*}

Figure~\ref{fig:moneyplots_SRmask} shows the \MET distributions in the monojet and mono-V signal regions.
The background prediction is obtained from a combined fit in all the control samples, excluding the signal region.
Data are found to be in agreement with the SM prediction.
Tables~\ref{tab:MonojetYields} and~\ref{tab:MonoVYields} show the estimated yields of background processes in the monojet and mono-V signal regions, respectively,
along with the observed event yields in the two signal regions. The correlations between the uncertainties across all the \MET bins in the two signal regions
are reported in Appendix~\ref{app:yields}. These results can be used with the simplified likelihood approach detailed in Ref.~\cite{CMS-NOTE-2017-001}
for reinterpretations in terms of models not studied in this paper.

\begin{table}[htb]
\topcaption{Expected event yields in each \MET bin for various background processes in the monojet signal region. The background yields and the
corresponding uncertainties are obtained after performing a combined fit to data in all the control samples, but excluding data in the signal region.
The observed event yields in the monojet signal region are also reported.}
\centering
\resizebox{\textwidth}{!}{
\begin{tabular}{c|c|c|c|c|c|c|c}
\hline
\MET [\GeVns{}] & \Zvvjets            & \Wlvjets             & Top quark           & Dibosons           & Other            & Total bkg. & Observed \\
\hline
200--230   & $71300 \pm 2200 $ & $ 54600 \pm 2300$ & $2140 \pm 320 $ & $ 1320 \pm 220$ & $2470 \pm 310 $ & $132100 \pm 4000\x $ & 140642 \\
230--260   & $ 39500 \pm 1300 $ & $ 27500 \pm 1200$ & $1060 \pm 160 $ & $ \x790\pm 130$ & $1090 \pm 130 $ & $ 69900 \pm 2200$ & 73114 \\
260--290   & $21900 \pm 670\x $ & $13600 \pm 550\x $ & $440 \pm 65  $ & $ 364 \pm 61 $ & $498 \pm  65 $ & $ 36800 \pm 1100$ & 38321 \\
290--320   & $12900 \pm 400\x $ & $ 7300 \pm 290 $ & $210 \pm 31  $ & $ 235 \pm 40 $ & $216 \pm 30  $ & $20780 \pm  630\x$ & 21417 \\
320--350   & $ 8000 \pm 280 $ & $ 4000 \pm 170 $ & $107 \pm 16  $ & $ 145 \pm 24 $ & $124 \pm 18  $ & $12340 \pm  400\x$ & 12525 \\
350--390   & $ 6100 \pm 220 $ & $ 2800 \pm 130 $ & $ \x74 \pm 11  $ & $ 111 \pm 19 $ & $ \x87 \pm 13  $ & $ 9160 \pm  320$ & 9515 \\
390--430   & $ 3500 \pm 160 $ & $1434 \pm 66\x  $ & $30.1 \pm 4.5 $ & $ 58.4 \pm 9.9$ & $33.4 \pm 5.3 $ & $ 5100 \pm  200$ & 5174 \\
430--470   & $2100 \pm 98\x  $ & $ 816 \pm 37  $ & $16.6 \pm 2.5 $ & $ 42.4 \pm 7.1$ & $16.3 \pm 2.7 $ & $ 3000 \pm  120$ & 2947 \\
470--510   & $1300 \pm 66\x  $ & $ 450 \pm 20  $ & $ \x7.4 \pm 1.1 $ & $ 24.6 \pm 4.1$ & $ \x9.6 \pm 1.6 $ & $1763 \pm   79\x$ & 1777 \\
510--550   & $ 735 \pm 39  $ & $ 266 \pm 13  $ & $ \x5.2 \pm 0.8 $ & $ 18.5 \pm 3.1$ & $ \x7.0 \pm 1.3 $ & $1032 \pm   48\x$ & 1021 \\
550--590   & $ 513 \pm 31  $ & $152 \pm 8\x   $ & $ \x2.4 \pm 0.4 $ & $ 13.5 \pm 2.3$ & $ \x1.1 \pm 0.3 $ & $ 683 \pm   37$ & 694 \\
590--640   & $ 419 \pm 23  $ & $120 \pm 6\x   $ & $ \x1.5 \pm 0.2 $ & $ 10.6 \pm 1.8$ & $ \x2.1 \pm 0.4 $ & $ 554 \pm   28$ & 554 \\
640--690   & $ 246 \pm 16  $ & $ 62.8 \pm 3.8 $ & $ \x1.3 \pm 0.2 $ & $ 11.4 \pm 1.9$ & $ \x1.0 \pm 0.2 $ & $ 322 \pm   19$ & 339 \\
690--740   & $ 139 \pm 11  $ & $ 34.2 \pm 2.4 $ & $ \x0.6 \pm 0.1 $ & $ \x4.2\pm 0.7$ & $ \x0.20 \pm 0.07$ & $ 178 \pm   13$ & 196 \\
740--790   & $ 97.2 \pm 7.2 $ & $ 22.7 \pm 1.7 $ & $ \x0.22 \pm 0.03$ & $ \x1.4\pm 0.2$ & $ \x0.63 \pm 0.12$ & $ 122 \pm    8\x$ & 123 \\
790--840   & $ 59.8 \pm 5.8 $ & $ 12.9 \pm 1.2 $ & $ \x0.13 \pm 0.02$ & $ \x1.5\pm 0.3$ & $ \x0.05 \pm 0.02$ & $ 74.5 \pm  6.6$ & 80 \\
840--900   & $ 64.3 \pm 6.4 $ & $ 12.3 \pm 1.1 $ & $ \x0.24 \pm 0.04$ & $ 0.92 \pm 0.1$ & $ \x0.03 \pm 0.01$ & $ 77.8 \pm  7.2$ & 68 \\
900--960   & $ 31.5 \pm 4.3 $ & $ \x6.0 \pm 0.7 $ & $ \x0.21 \pm 0.03$ & $ 0.74 \pm 0.1$ & $ \x0.01 \pm 0.01$ & $ 38.4 \pm  4.8$ & 37 \\
\x960--1020  & $ 20.8 \pm 3.0 $ & $ \x3.4 \pm 0.5 $ &  \x---             & $ 0.94 \pm 0.2$ & $ \x0.01 \pm 0.01$ & $ 25.1 \pm  3.4$ & 23 \\
1020--1090 & $ 16.3 \pm 2.6 $ & $ \x3.1 \pm 0.5 $ & $ \x0.04 \pm 0.01$ & $ \x1.6\pm 0.3$ & $ \x0.01 \pm 0.01$ & $ 21.1 \pm  3.0$ & 12 \\
1090--1160 & $ \x8.1 \pm 1.8 $ & $  \x1.3 \pm 0.3 $ & \x---            &   \x---          &  \x ---             & $ \x9.4 \pm  1.9$ & 7 \\
$ {>}1160$  & $ 18.6 \pm 2.7 $ & $ \x2.7 \pm 0.4 $ &  \x---            & $  \x1.3\pm 0.2$ &  \x---             & $ 22.6 \pm  3.0$ & 26 \\
\hline
\end{tabular}
}

\label{tab:MonojetYields}
\end{table}

\begin{table}[!htb]
\topcaption{Expected event yields in each \MET bin for various background processes in the mono-V signal region. The background yields and the
corresponding uncertainties are obtained after performing a combined fit to data in all the control samples, excluding data in the signal region.
The observed event yields in the mono-V signal region are also reported.}
\centering
\resizebox{\textwidth}{!}{
\begin{tabular}{c|c|c|c|c|c|c|c}
\hline
\MET [\GeVns{}] & \Zvvjets            & \Wlvjets             & Top quark       & Dibosons            & Other               & Total bkg.         & Observed \\
\hline
250--300   & $1700 \pm  88\x $ & $    1100 \pm  65\x $ & $ 171 \pm 24 $ & $ 195 \pm 35  $ & $ \x49.4 \pm 10.8$ & $ 3220 \pm 130$ & 3395 \\
300--350   & $1180 \pm  68\x $ & $ 627 \pm  37 $ & $ 68.8 \pm 9.7$ & $ 135 \pm 24  $ & $    44.2 \pm 7.2 $ & $    2050 \pm 88\x $ & 2162 \\
350--400   & $ 629 \pm  37 $ & $ 314 \pm  21 $ & $ 28.9 \pm 4.1$ & $  68.5 \pm 12\phantom{.}  $ & $ \x8.0 \pm 1.8 $ & $    1048 \pm 51\x $ & 1093 \\
400--500   & $ 500 \pm  33 $ & $ 181 \pm  13 $ & $ 21.4 \pm 3.0$ & $  62.8 \pm 11\phantom{.}  $ & $    10.1 \pm 1.8 $ & $ 775 \pm 40 $ & 780  \\
500--600   & $ 131 \pm  12 $ & $ 38.5 \pm  3.4$ & $ \x2.9\pm 0.4$ & $ 16.8 \pm 3.0 $ & $ \x4.1 \pm 0.8 $ & $ 193 \pm 14 $ & 207  \\
600--750   & $ 57.1 \pm  5.9$ & $ 15.6 \pm  1.6$ & $ \x1.0\pm 0.1$ & $ \x9.8\pm 1.7 $ & $ \x0.8 \pm 0.1 $ & $ 84.2 \pm 6.9$ & 90   \\
$ {>}750$   & $ 16.5 \pm  2.7$ & $ \x3.6 \pm  0.6$ &         \x---       & $ \x4.7\pm 0.8 $ & $ \x0.01 \pm 0.01$ & $ 24.8 \pm 3.1$ & 27   \\
\hline
\end{tabular}
}

\label{tab:MonoVYields}
\end{table}

\begin{figure*}[!htb]
\centering
\includegraphics[width=0.46\textwidth]{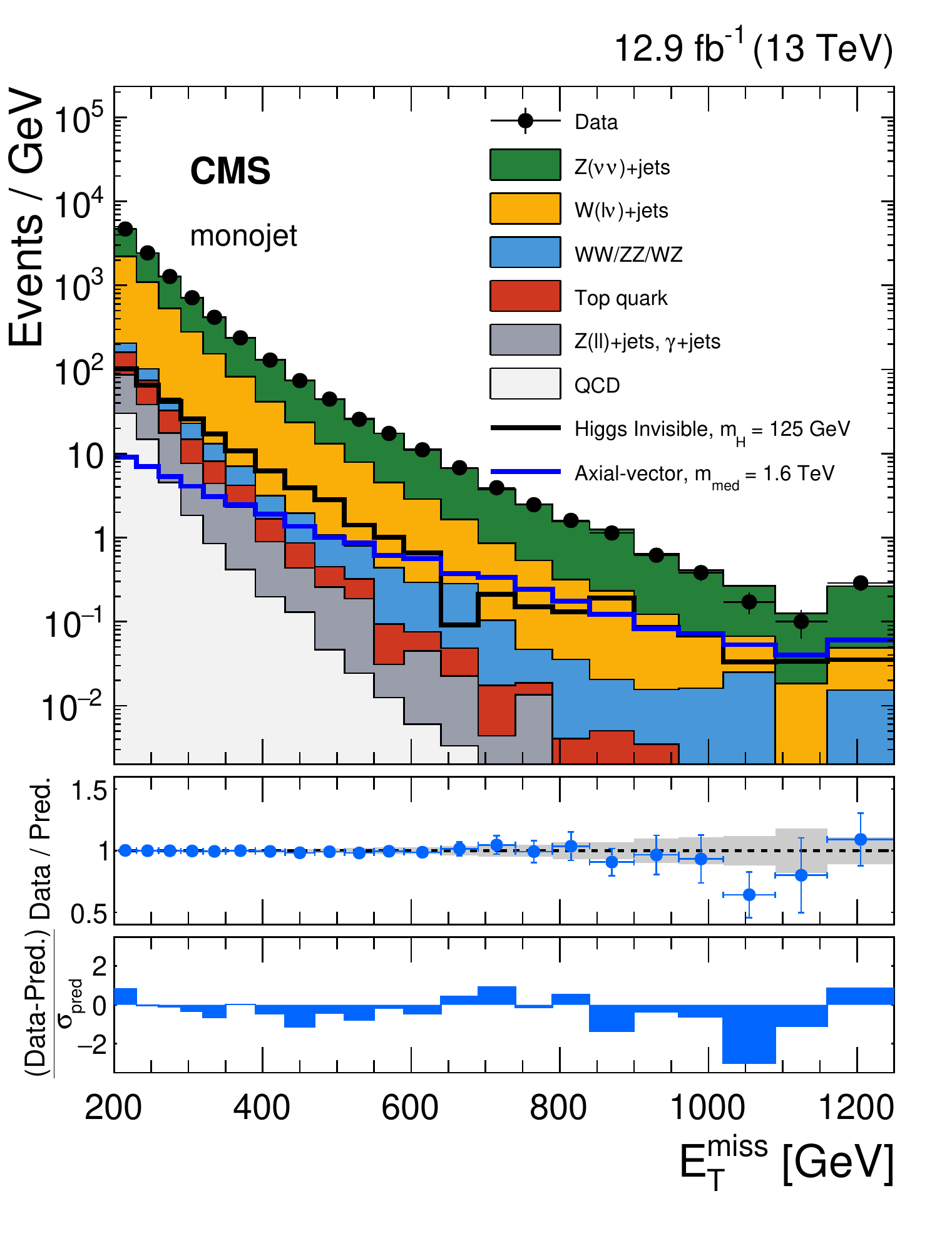}
\includegraphics[width=0.46\textwidth]{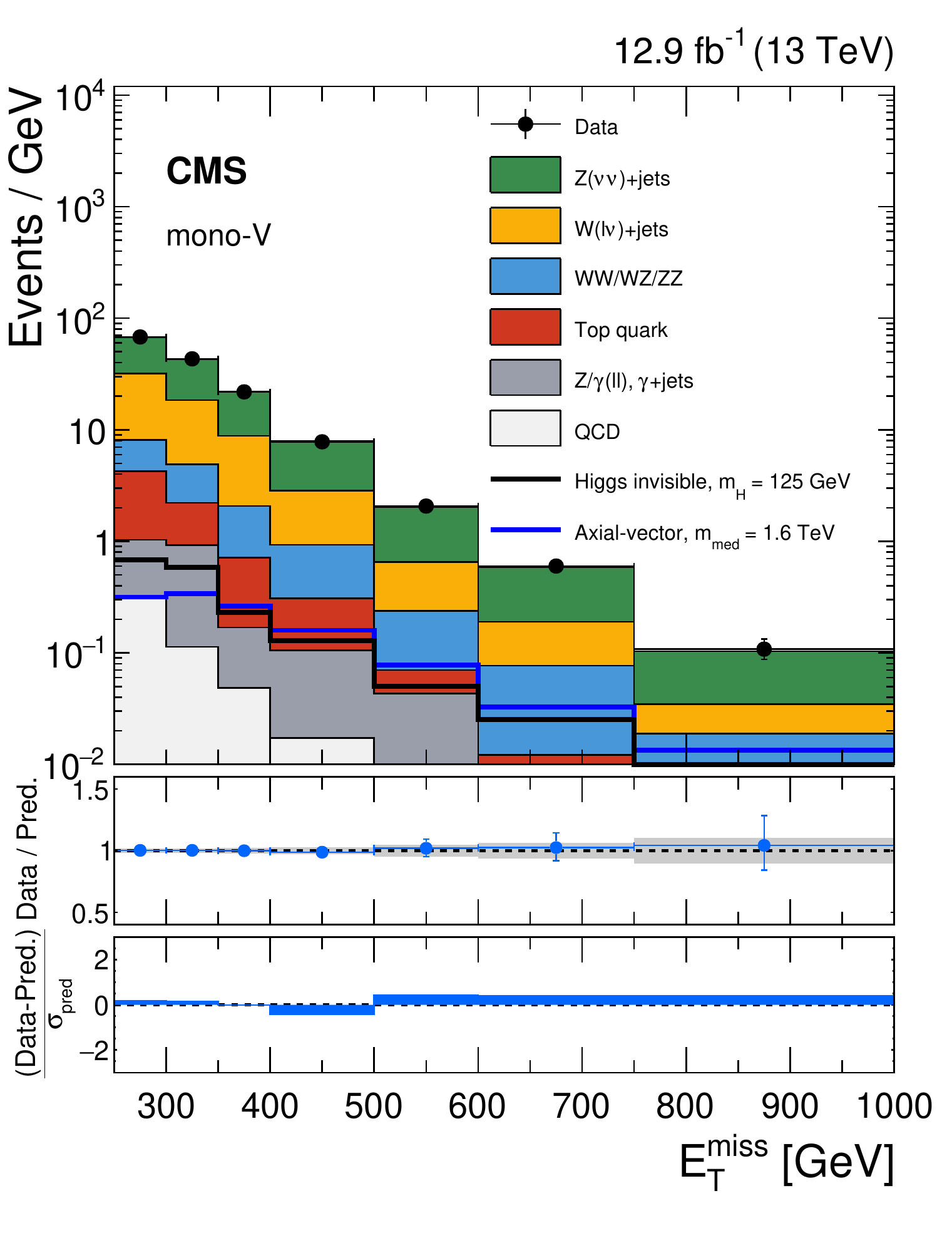}
\caption{
Observed \MET distribution in the monojet (left) and mono-V (right) signal regions compared with the background expectations for various SM processes
evaluated after performing a combined fit to the data in all the control samples, as well as in the signal region. The fit is performed assuming the absence of any signal.
The last bin includes all events with $\MET > $ 1160 (750)\GeV for the monojet (mono-V) category. Expected signal distributions for a 125\GeV Higgs boson
decaying exclusively to invisible particles, and for a 1.6\TeV axial-vector mediator decaying to 1\GeV DM particles, are overlaid.
The ratio of data and the post-fit background prediction is shown for both the monojet and mono-V signal regions.
The gray bands in these ratio plots indicate the post-fit uncertainty in the background prediction.
Finally, the distributions of the pulls, defined as the difference between data and the post-fit background prediction relative to the post-fit uncertainty in the prediction,
are also shown in the lower panels.
}
\label{fig:moneyplots}

\end{figure*}

Figure~\ref{fig:moneyplots} shows the \MET distributions where the background estimates
have been computed after including events from the signal region in the fit, but assuming the absence of any signal. The comparison of this fit with an alternative fit assuming
the presence of signal is used to set limits on the DM signal cross section.

\subsection{Dark matter interpretation}
The results of the search are interpreted in terms of simplified DM models for the monojet and mono-V final states, assuming a vector, axial-vector,
scalar, or pseudoscalar mediator decaying into a pair of fermionic DM particles.
These results supersede those from the earlier CMS publications in the same final states~\cite{Khachatryan:2014rra,Khachatryan:2016mdm}.

The mediators are assumed to interact with the pair of DM particles with coupling strength $\mathrm{g}_{\mathrm{DM}} = 1$.
The spin-1 mediators are assumed to interact with SM quarks with coupling strength $\mathrm{g}_{\mathrm{q}} = 0.25$.
The spin-0 mediators are assumed to couple to the quarks through SM-like Yukawa interactions with the coupling strength modifier $\mathrm{g}_{\mathrm{q}} = 1$.
The width of the mediators is determined assuming they interact only with the SM particles and the DM particle.
The choice of all the signal model parameters follows the recommendations from Ref.~\cite{Abercrombie:2015wmb} (Sec. 2.1 and 2.2).
Uncertainties of 20 and 30\% are assigned to the inclusive signal cross section in the case of the spin-1 and spin-0 mediators, respectively.
These include the renormalization and factorization scale uncertainties, and the PDF uncertainty.

\begin{figure*}[hbtp]
\centering
\includegraphics[width=0.49\textwidth]{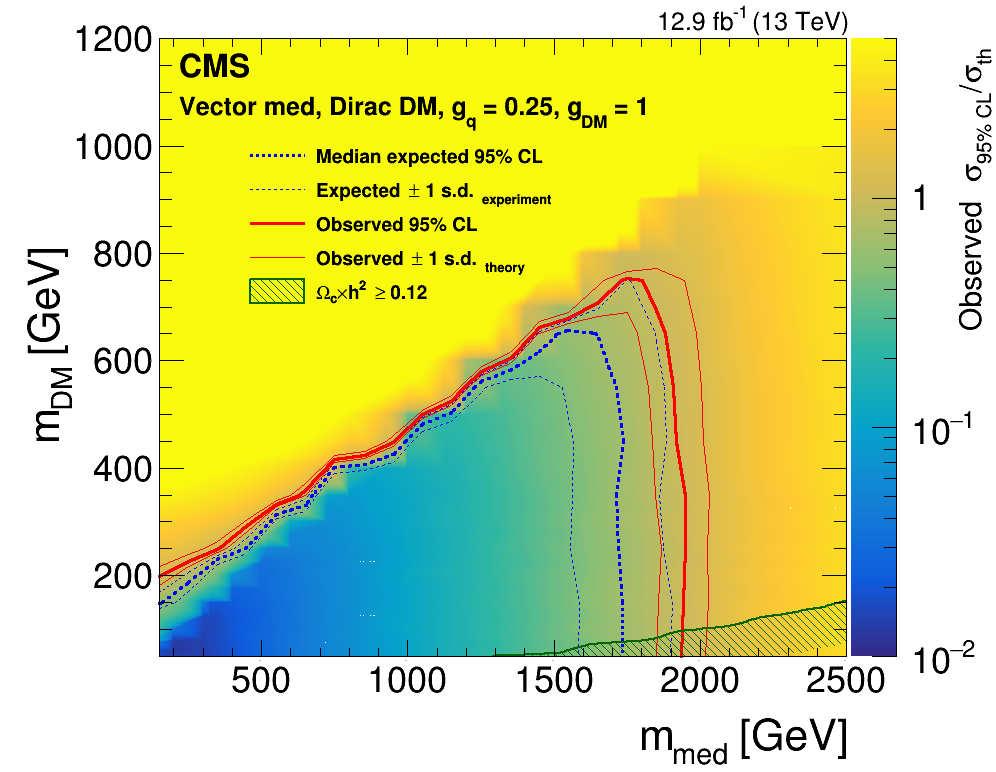}
\includegraphics[width=0.49\textwidth]{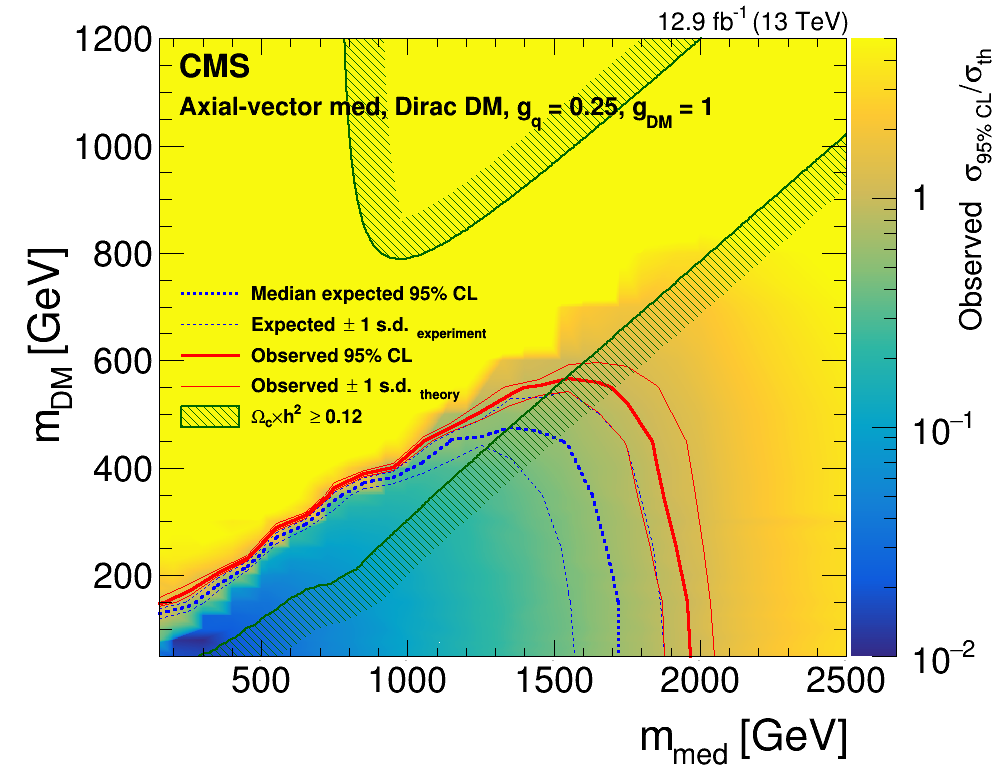}
\caption{
Exclusion limits at 95\% CL on the signal strength $\mu=\sigma/\sigma_{\text{th}}$ in the $m_{\text{med}}$--$m_{\text{DM}}$ plane assuming vector (left) and axial-vector (right)
mediators. The limits are shown for $m_{\text{med}}$ between 150\GeV and 2.5\TeV, and $m_{\text{DM}}$ between 50\GeV and 1.2\TeV.
While the excluded area is expected to extend below these minimum values of $m_{\text{med}}$ and $m_{\text{DM}}$,
the axes do not extend below these values as the signal simulation was not performed in this region.
The solid (dotted) red (blue) line shows the contour for the observed (expected) exclusion. The solid contours around the observed limit and the dashed contours
around the expected limit represent one standard deviation theoretical uncertainties in the signal cross section and the combination of the statistical
and experimental systematic uncertainties, respectively. Constraints from the Planck satellite experiment~\cite{Ade:2015xua} are shown with the dark green contours and associated hatching.
The hatched area indicates the region where the DM density exceeds the observed value.
}
\label{fig:scan_spin1}

\end{figure*}

\begin{figure*}[!h]
\centering
\includegraphics[width=0.49\textwidth]{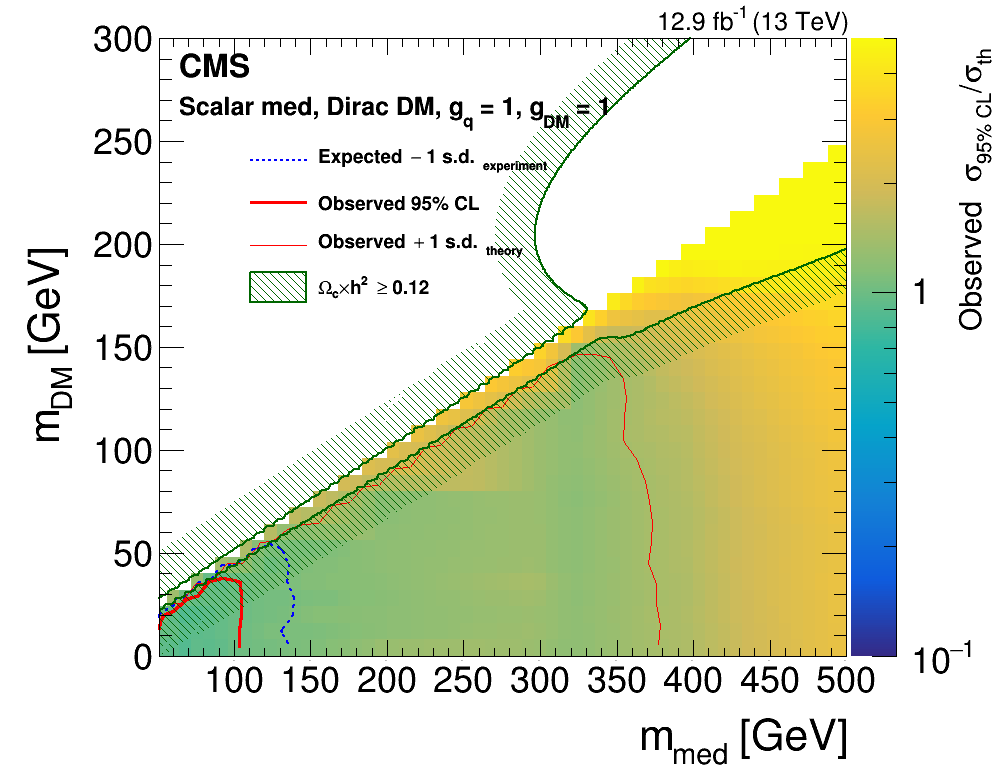}
\includegraphics[width=0.49\textwidth]{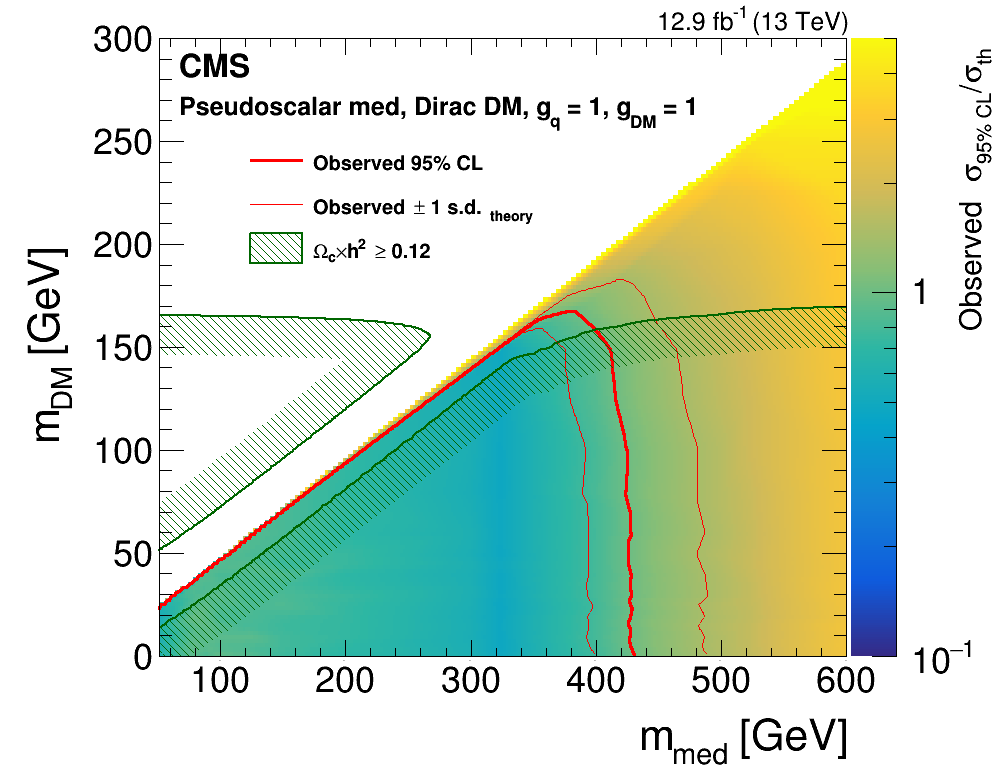}
\caption{
Exclusion limits at 95\% CL on signal strength the $\mu=\sigma/\sigma_{\text{th}}$ in the $m_{\text{med}}$--$m_{\text{DM}}$ plane assuming scalar (left) and pseudoscalar (right)
mediators. The limits are shown for $m_{\text{med}}$ between 50 and 500\GeV, and $m_{\text{DM}}$ between 0 and 300\GeV.
While the excluded area is expected to extend below the minimum value of $m_{\text{med}}$,
the axis does not extend below this value as the signal simulation was not performed in this region.
The red line shows the contour for the observed  exclusion. The solid red contours around the observed limit
represent one standard deviation theoretical uncertainties in the signal cross section.
The dashed blue contour in the case of the scalar mediator shows the $-1\sigma$ deviation due to the combination of the statistical and experimental systematic uncertainties.
Constraints from the Planck satellite experiment~\cite{Ade:2015xua} are shown with the dark green contours and associated hatching.
The hatched area indicates the region where the DM density exceeds the observed value.
}
\label{fig:scan_spin0}

\end{figure*}

\begin{figure*}[hbtp]
\centering
\includegraphics[width=0.45\textwidth]{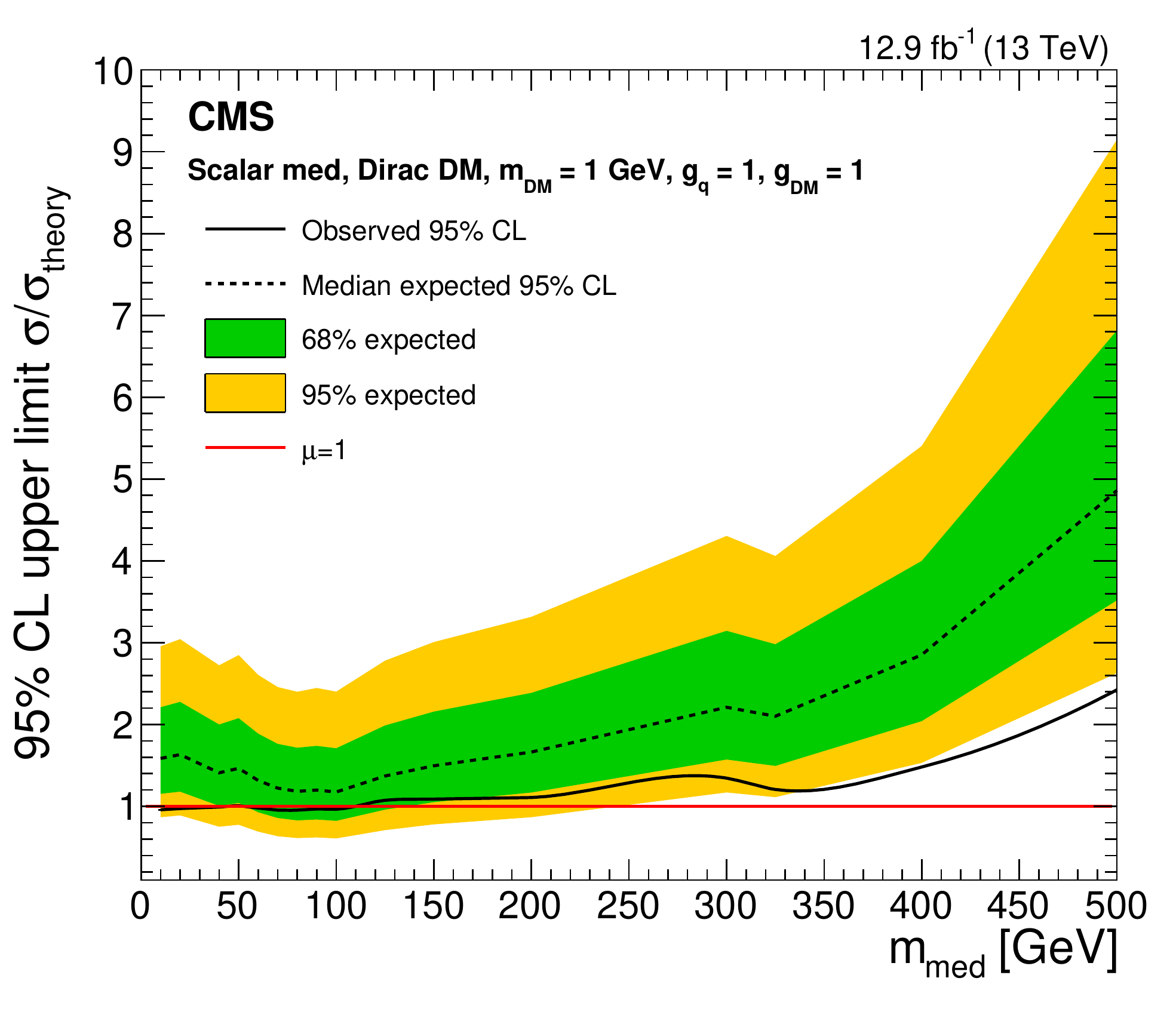}
\includegraphics[width=0.45\textwidth]{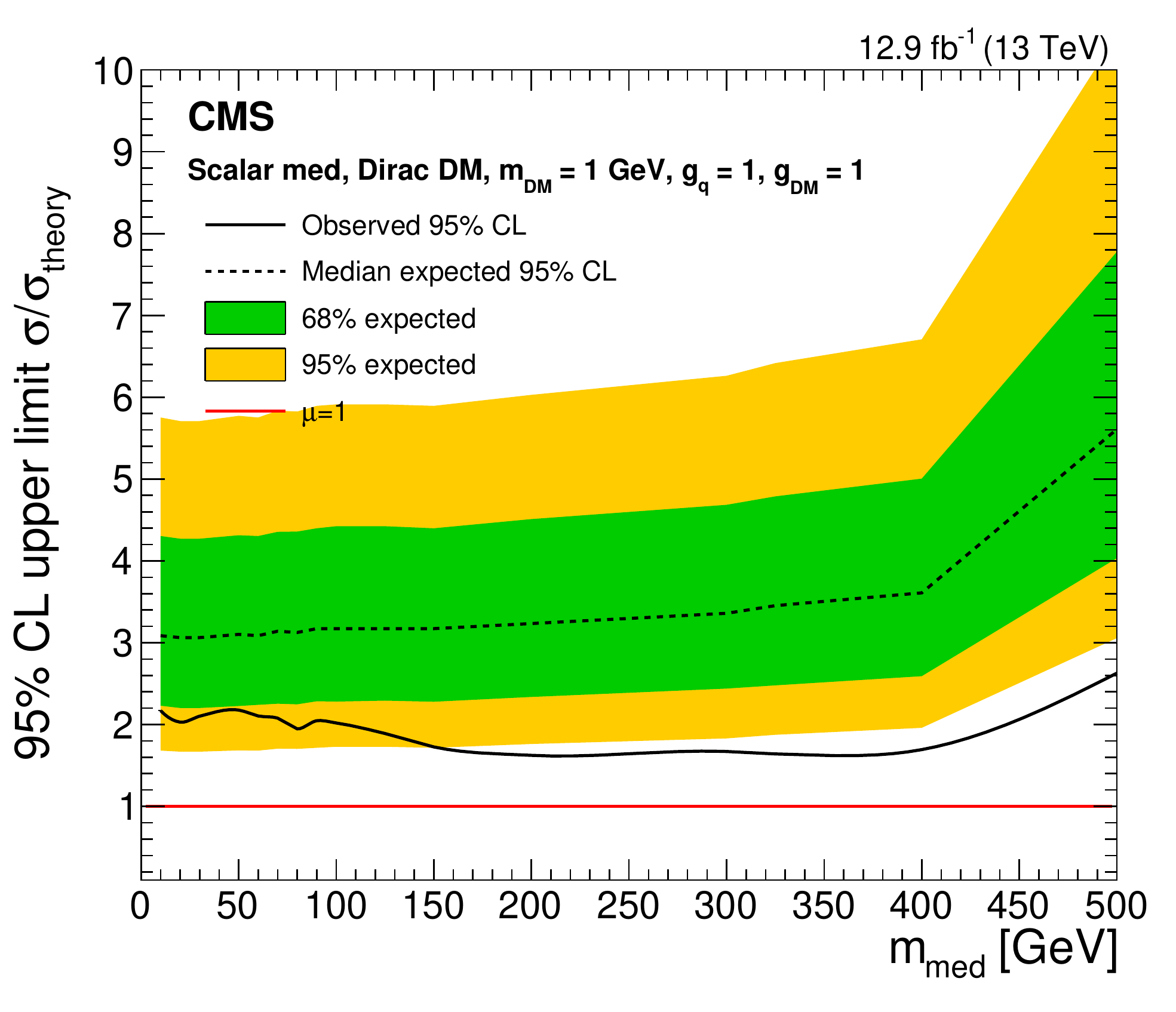}
\includegraphics[width=0.45\textwidth]{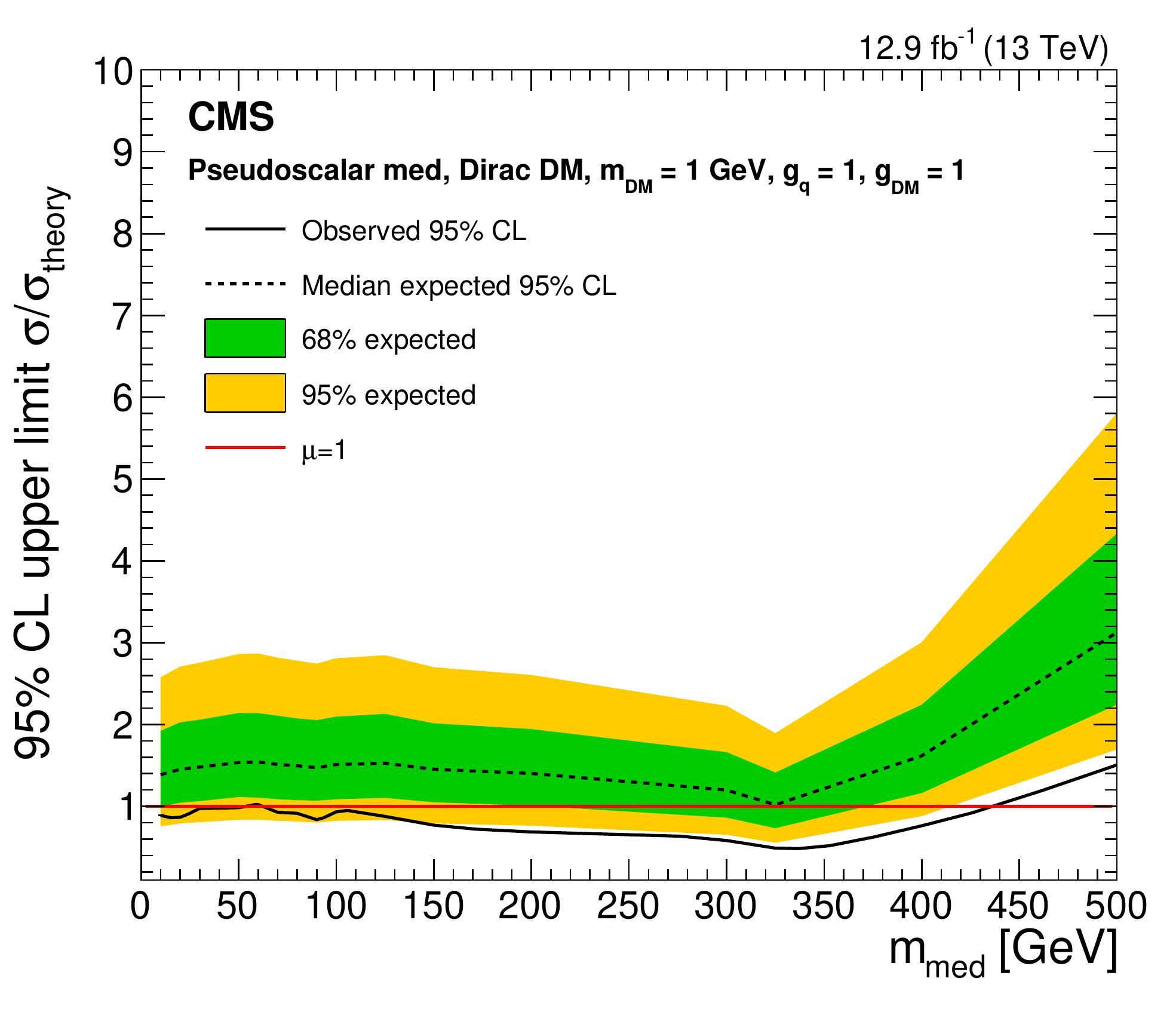}
\caption{
Expected (dotted black line) and observed (solid black line) 95\% CL upper limits on the signal strength $\mu$ as a function of the mediator mass for the spin-0 models.
The horizontal red line denotes $\mu = 1$. Limits for the scalar model on the combined cross section of the monojet and mono-V processes (upper left).
Limits for the scalar (upper right) and pseudoscalar (bottom) models, respectively, assuming only the monojet signal process.
}
\label{fig:1d}

\end{figure*}

Upper limits are computed at 95\% CL on the ratio of the signal cross section to the predicted cross section, denoted by
$\mu=\sigma/\sigma_{\text{th}}$, with the CL$_{\mathrm{s}}$ method~\cite{Junk:1999kv,Read:2002av}, using the asymptotic approximation~\cite{Cowan:2010js}.
Limits are obtained as a function of the mediator mass, $m_{\text{med}}$, and the DM mass, $m_{\text{DM}}$.
In the case of the vector, axial-vector and scalar mediators, limits are computed on the combined cross section due to the monojet and mono-V signal processes.
In the case of the pseudoscalar mediator, limits are computed assuming only the monojet signal process.
The mono-V signal process (Fig.~\ref{fig:Spin0_FD}, right), in which a pseudoscalar mediator couples directly to vector bosons, is ill-defined
without making additional assumptions~\cite{Gunion:1990kf} and therefore is not included.
Figure~\ref{fig:scan_spin1} shows the exclusion contours in the $m_{\text{med}}$--$m_{\text{DM}}$ plane for the vector and axial-vector mediators.
Mediator masses up to 1.95 \TeV and DM masses up to 750 and 550\GeV are excluded for the vector and axial-vector models, respectively, at 95\% CL.
Figure~\ref{fig:scan_spin0} shows the exclusion contours in the $m_{\text{med}}$--$m_{\text{DM}}$ plane for the scalar and pseudoscalar mediators.
For scalar mediators, masses up to 100\GeV and DM masses up to 35\GeV are excluded at 95\% CL, and no exclusion is expected or observed considering only the monojet signal process.
Pseudoscalar mediator masses up to 430\GeV and DM masses up to 170\GeV are excluded at 95\% CL.
Figure~\ref{fig:1d} shows the limits for the spin-0 models as a function of the mediator mass, assuming the DM mass to be 1\GeV.
In the case of the scalar mediator limits are computed for the monojet signal process, and for the combination of the monojet and mono-V signal processes.

Figures~\ref{fig:scan_spin1} and \ref{fig:scan_spin0} also show the constraints from the observed cosmological relic density of DM as determined from measurements of the
cosmic microwave background by the Planck satellite experiment~\cite{Ade:2015xua}. The expected DM abundance is estimated using the thermal freeze-out mechanism
implemented in the {\sc MadDM}~\cite{Backovic:2013dpa} package, and compared to the observed cold DM density $\Omega_c h^2=0.12$~\cite{Ade:2013zuv}, where $\Omega_c$ is the DM
relic abundance and $h$ is the Hubble constant, under the assumption that a single DM  particle describes DM interactions in the early universe and that this particle
only interacts with SM particles through the considered simplified model~\cite{Pree:2016hwc,Backovic:2015soa}.

\begin{figure*}[hbtp]
\centering
\includegraphics[width=0.45\textwidth]{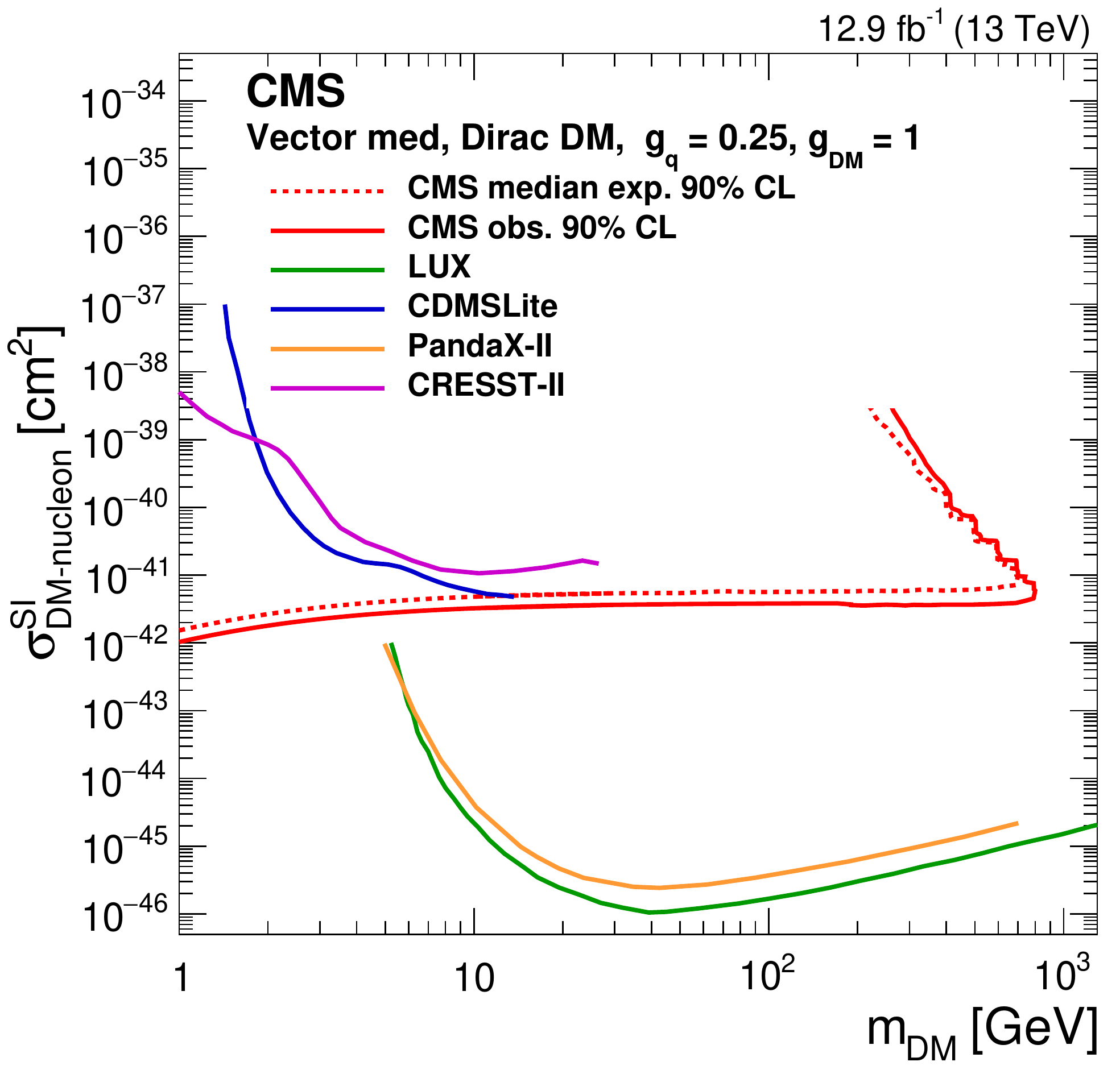}
\includegraphics[width=0.45\textwidth]{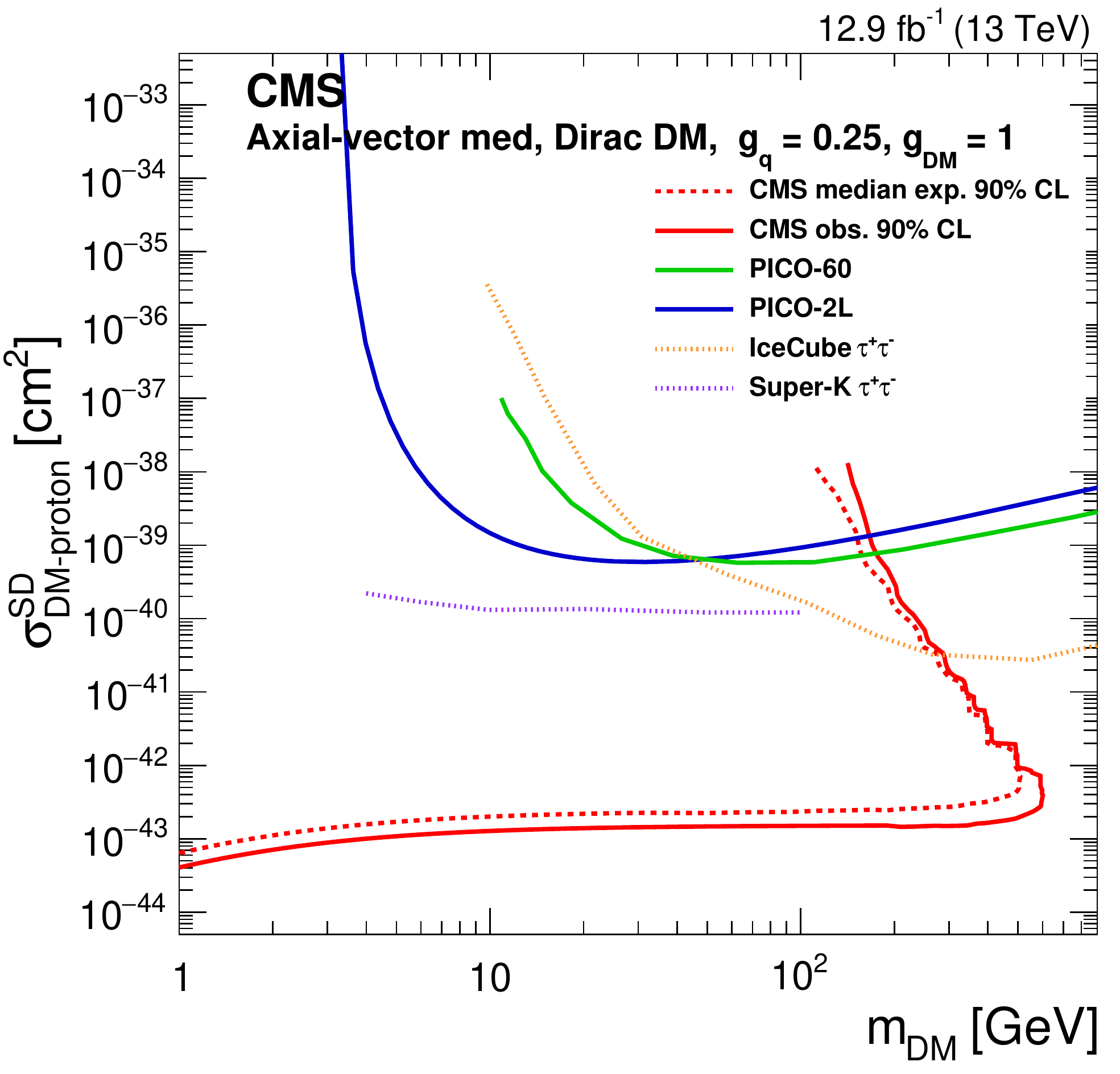}
\caption{
Exclusion limits at 90\% CL in the $m_{\text{DM}}$ vs. $\sigma_{\text{SI/SD}}$ plane for vector (left) and axial-vector (right) mediator models.
The solid (dotted) red line shows the contour for the observed (expected) exclusion in this search.
Limits from the CDMSLite~\cite{Agnese:2015nto}, LUX~\cite{Akerib:2016vxi}, PandaX-II~\cite{Tan:2016zwf}, and CRESST-II~\cite{cresst} experiments are shown for the vector mediator.
Limits from the PICO-2L~\cite{Amole:2016pye}, PICO-60~\cite{Amole:2015pla}, IceCube~\cite{Aartsen:2016exj}, and Super-Kamiokande \cite{Choi:2015ara}
experiments are shown for the axial-vector mediator.}
\label{fig:nucleon}
\end{figure*}

\begin{figure*}[hbtp]
\centering
\includegraphics[width=0.45\textwidth]{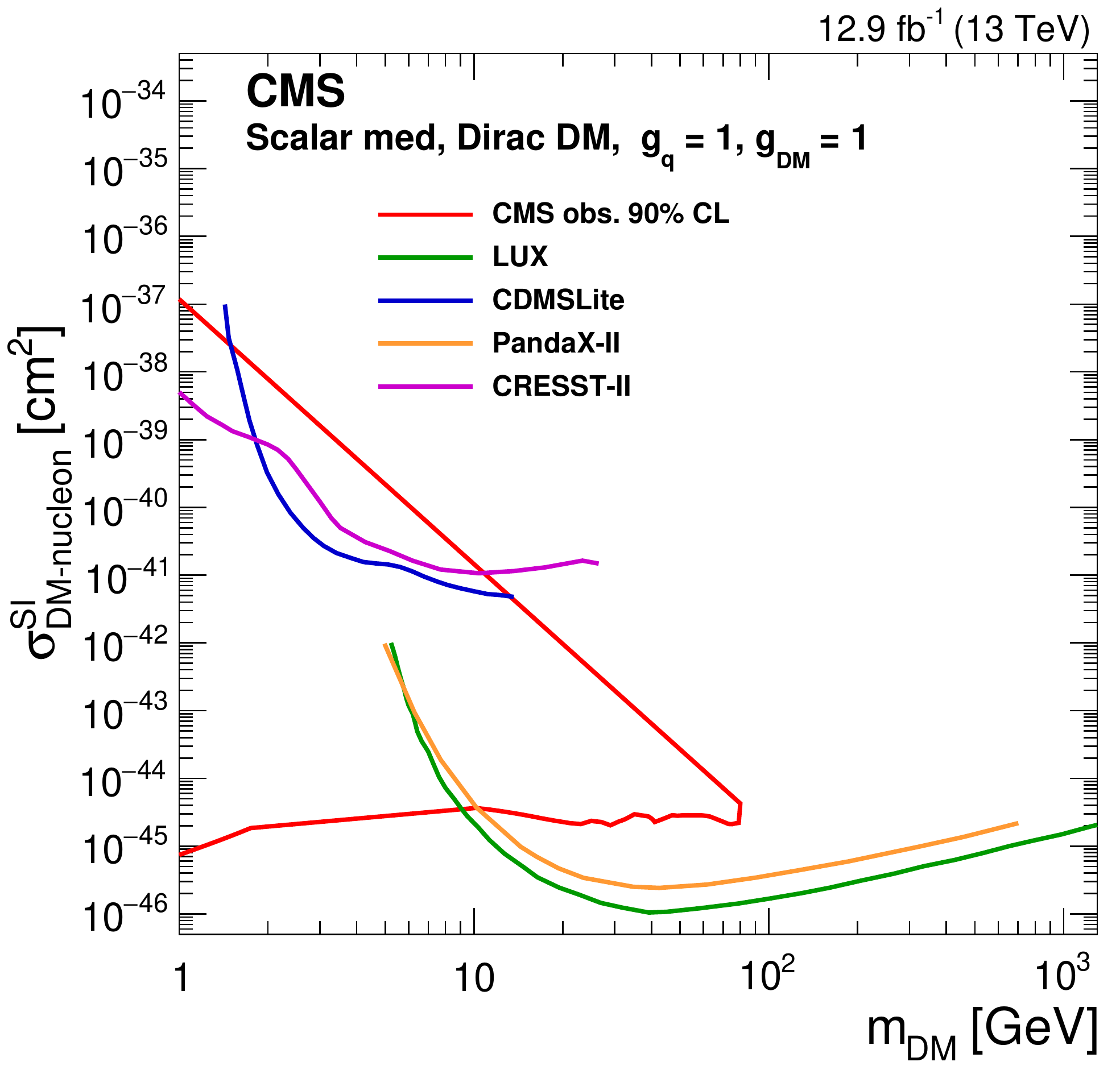}
\includegraphics[width=0.45\textwidth]{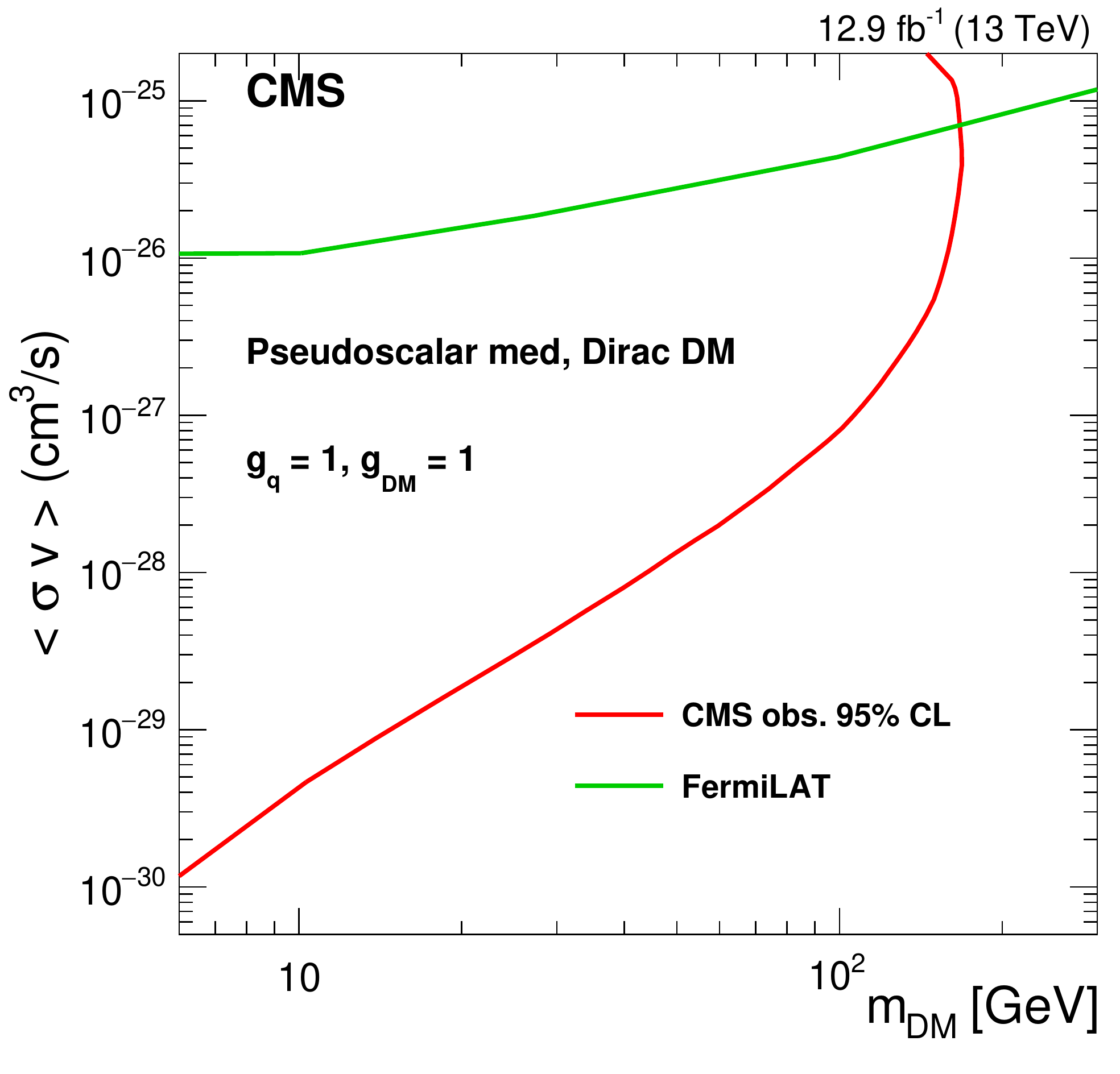}
\caption{
Exclusion limits at 90\% CL in the $m_{\text{DM}}$ vs. $\sigma_{\text{SI/SD}}$ plane for the scalar mediator model (left).
The observed exclusion in this search (red line) is compared to the results from the
CDMSLite~\cite{Agnese:2015nto}, LUX~\cite{Akerib:2016vxi}, PandaX-II~\cite{Tan:2016zwf}, and CRESST-II~\cite{cresst} experiments.
For the pseudoscalar mediator (right), limits at 95\% CL are compared to the velocity-averaged DM annihilation cross section upper limits from Fermi--LAT~\cite{Ackermann:2015zua}.
There are no comparable limits from direct detection experiments as the scattering cross section between DM particles and SM quarks is suppressed at nonrelativistic
velocities for a pseudoscalar mediator~\cite{Haisch:2012kf,Berlin:2014tja}.
}
\label{fig:nucleon2}
\end{figure*}

The limits obtained using the simplified DM models may be compared to the results from direct and indirect DM detection experiments, which are usually expressed as
90\% CL upper limits on the DM-nucleon scattering cross sections. The approach outlined in Refs.~\cite{Buchmueller:2014yoa,Harris:2015kda,Boveia:2016mrp} is used to translate
the exclusion contours into the $m_{\text{DM}}$ vs. $\sigma_{\text{SI/SD}}$ plane where $\sigma_{\text{SI/SD}}$ are the spin-independent/spin-dependent DM-nucleon
scattering cross sections. These limits are shown in Fig.~\ref{fig:nucleon} for the vector and axial-vector mediators, and in Fig.~\ref{fig:nucleon2} (left) for the scalar mediator.
For the scalar mediator model, only the contributions from heavy quarks (charm, bottom, and top) are taken into account while evaluating the limit on the
DM-nucleon cross section, as done in Ref.~\cite{Khachatryan:2016mdm}.
When compared to the results from direct detection experiments, the limits obtained from this search provide stronger constraints for dark matter masses less than 5, 9, and 550\GeV,
assuming vector, scalar, and axial-vector mediators, respectively.
In the case of the pseudoscalar mediator, the 95\% CL upper limits are compared in Fig.~\ref{fig:nucleon2} (right) with the indirect detection results in terms of the
velocity-averaged DM annihilation cross section from the Fermi--LAT Collaboration~\cite{Ackermann:2015zua}, and provide stronger constraints for DM masses less than 200\GeV.

\subsection{Invisible decays of the Higgs boson}
\begin{figure*}[hbtp]\centering
\includegraphics[width=0.5\textwidth]{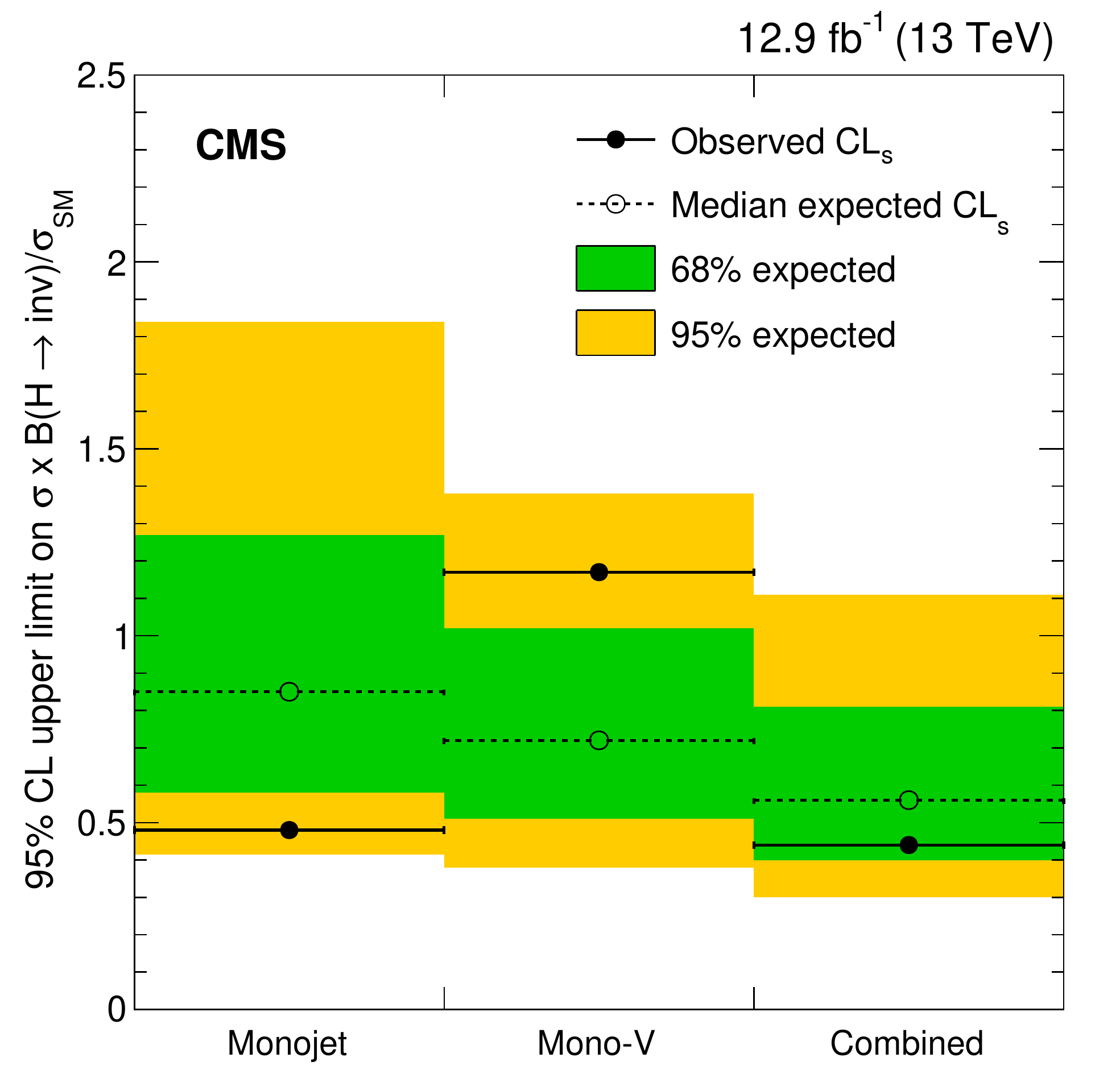}
\caption{Expected (dotted black line) and observed (solid black line) 95\% CL upper limits on the invisible branching fraction of a 125\GeV SM-like Higgs boson. Limits are shown for the monojet and mono-V categories separately, and also for their combination.}
\label{fig:HinvLimitsPlot}\end{figure*}

The results of this search are also interpreted in terms of an upper limit on the product of the cross section and branching fraction \brhinv, relative to the
predicted cross section ($\sigma_{\text{SM}}$) of the Higgs boson assuming SM interactions,
where the Higgs boson is produced through gluon fusion (ggH) along with a jet; in association with a vector boson (ZH, WH);
or through vector boson fusion (VBF). The predictions for the Higgs boson production cross section
and the corresponding theoretical uncertainties are taken from the recommendations of the LHC Higgs cross section working group~\cite{deFlorian:2016spz}.
If the production cross section of the Higgs boson is assumed to be the same as $\sigma_{\text{SM}}$, this limit can be used to constrain the invisible branching fraction of the Higgs boson.
The observed (expected) 95\% CL upper limit on the invisible branching fraction of the Higgs boson, $\sigma \brhinv / \sigma_{\text{SM}}$,
is found to be 0.44 (0.56). The limits are summarized in Fig.~\ref{fig:HinvLimitsPlot}. Table~\ref{tab:HiggsLimits} shows the individual limits for the monojet and mono-V categories.
While these limits on \brhinv~are not as strong as the combined ones from Ref.~\cite{Khachatryan:2016whc}, they are obtained from an independent data sample and
therefore will contribute to future combinations.

\begin{table}[htb]
\topcaption{Expected and observed 95\% CL upper limits on the invisible branching fraction of the Higgs boson.
Limits are tabulated for the monojet and mono-V categories separately, and for their combination. The one standard deviation uncertainty range on the expected limits is listed.
The signal composition in terms of gluon fusion, vector boson fusion, and an associated production with a W or Z boson is also provided.}
\label{tab:HiggsLimits}
\centering
\resizebox{\textwidth}{!}{
\begin{tabular}{l|c|c|c|c} \hline
\multirow{2}{*}{Category}          & Expected     & Observed     & \multirow{2}{*}{${\pm} 1$ s.d.}         & Expected signal \\
& limit        & limit        &                      & composition \\
\hline
Mono-V            & 0.72         & 1.17         & [0.51--1.02]         & 39.6\% ggH, \x6.9\% VBF, 32.4\% WH, 21.1\% ZH \\
Monojet           & 0.85         & 0.48         & [0.58--1.27]         & 71.5\% ggH, 20.3\% VBF, \x4.4\% WH, \x3.8\% ZH \\
Combined          & 0.56         & 0.44         & [0.40--0.81]         & --- \\
\hline
\end{tabular}
}
\end{table}

\section{Summary}
A search for dark matter (DM) is presented using events with jets and large missing transverse momentum in a $\sqrt{s} = 13\TeV$
proton-proton collision data set corresponding to an integrated luminosity of 12.9\fbinv.
The search also exploits events with a hadronic decay of a W or Z boson reconstructed as a single large-radius jet. No significant excess is observed with respect to the standard model
backgrounds. Limits are computed on the DM production cross section using simplified models in which DM production is mediated by spin-1 or spin-0 particles.
Vector and axial-vector mediators with masses up to 1.95 \TeV are excluded at 95\% confidence level, assuming a coupling strength of 0.25 between the mediators and the standard model fermions, and
a coupling strength of 1.0 between the mediators and the DM particles.
The results of this search provide the strongest constraints on DM pair production through vector and axial-vector mediators at a particle collider.
Scalar and pseudoscalar mediators with masses up to 100 and 430\GeV, respectively, are excluded at 95\% confidence level, assuming the coupling of the spin-0 mediators with DM particles
to be 1.0 and the coupling of the spin-0 mediators with standard model fermions to be the same as the standard model Yukawa interactions.
When compared to the direct detection experiments, the limits obtained from this search provide stronger constraints for dark matter masses less than 5, 9, and 550\GeV,
assuming vector, scalar, and axial-vector mediators, respectively. The search yields stronger constraints for dark matter masses less than 200\GeV, assuming
a pseudoscalar mediator, when compared to the indirect detection results from Fermi--LAT.
The search also yields an observed (expected) 95\% confidence level upper limit of 0.44 (0.56) on the invisible branching fraction of a standard model-like 125\GeV Higgs boson,
assuming the standard model production cross section.

\begin{acknowledgments}

\hyphenation{Bundes-ministerium Forschungs-gemeinschaft Forschungs-zentren Rachada-pisek} We congratulate our colleagues in the CERN accelerator departments for the excellent performance of the LHC and thank the technical and administrative staffs at CERN and at other CMS institutes for their contributions to the success of the CMS effort. In addition, we gratefully acknowledge the computing centres and personnel of the Worldwide LHC Computing Grid for delivering so effectively the computing infrastructure essential to our analyses. Finally, we acknowledge the enduring support for the construction and operation of the LHC and the CMS detector provided by the following funding agencies: the Austrian Federal Ministry of Science, Research and Economy and the Austrian Science Fund; the Belgian Fonds de la Recherche Scientifique, and Fonds voor Wetenschappelijk Onderzoek; the Brazilian Funding Agencies (CNPq, CAPES, FAPERJ, and FAPESP); the Bulgarian Ministry of Education and Science; CERN; the Chinese Academy of Sciences, Ministry of Science and Technology, and National Natural Science Foundation of China; the Colombian Funding Agency (COLCIENCIAS); the Croatian Ministry of Science, Education and Sport, and the Croatian Science Foundation; the Research Promotion Foundation, Cyprus; the Secretariat for Higher Education, Science, Technology and Innovation, Ecuador; the Ministry of Education and Research, Estonian Research Council via IUT23-4 and IUT23-6 and European Regional Development Fund, Estonia; the Academy of Finland, Finnish Ministry of Education and Culture, and Helsinki Institute of Physics; the Institut National de Physique Nucl\'eaire et de Physique des Particules~/~CNRS, and Commissariat \`a l'\'Energie Atomique et aux \'Energies Alternatives~/~CEA, France; the Bundesministerium f\"ur Bildung und Forschung, Deutsche Forschungsgemeinschaft, and Helmholtz-Gemeinschaft Deutscher Forschungszentren, Germany; the General Secretariat for Research and Technology, Greece; the National Scientific Research Foundation, and National Innovation Office, Hungary; the Department of Atomic Energy and the Department of Science and Technology, India; the Institute for Studies in Theoretical Physics and Mathematics, Iran; the Science Foundation, Ireland; the Istituto Nazionale di Fisica Nucleare, Italy; the Ministry of Science, ICT and Future Planning, and National Research Foundation (NRF), Republic of Korea; the Lithuanian Academy of Sciences; the Ministry of Education, and University of Malaya (Malaysia); the Mexican Funding Agencies (BUAP, CINVESTAV, CONACYT, LNS, SEP, and UASLP-FAI); the Ministry of Business, Innovation and Employment, New Zealand; the Pakistan Atomic Energy Commission; the Ministry of Science and Higher Education and the National Science Centre, Poland; the Funda\c{c}\~ao para a Ci\^encia e a Tecnologia, Portugal; JINR, Dubna; the Ministry of Education and Science of the Russian Federation, the Federal Agency of Atomic Energy of the Russian Federation, Russian Academy of Sciences, the Russian Foundation for Basic Research and the Russian Competitiveness Program of NRNU ``MEPhI''; the Ministry of Education, Science and Technological Development of Serbia; the Secretar\'{\i}a de Estado de Investigaci\'on, Desarrollo e Innovaci\'on, Programa Consolider-Ingenio 2010, Plan de Ciencia, Tecnolog\'{i}a e Innovaci\'on 2013-2017 del Principado de Asturias and Fondo Europeo de Desarrollo Regional, Spain; the Swiss Funding Agencies (ETH Board, ETH Zurich, PSI, SNF, UniZH, Canton Zurich, and SER); the Ministry of Science and Technology, Taipei; the Thailand Center of Excellence in Physics, the Institute for the Promotion of Teaching Science and Technology of Thailand, Special Task Force for Activating Research and the National Science and Technology Development Agency of Thailand; the Scientific and Technical Research Council of Turkey, and Turkish Atomic Energy Authority; the National Academy of Sciences of Ukraine, and State Fund for Fundamental Researches, Ukraine; the Science and Technology Facilities Council, UK; the US Department of Energy, and the US National Science Foundation.

Individuals have received support from the Marie-Curie programme and the European Research Council and EPLANET (European Union); the Leventis Foundation; the A. P. Sloan Foundation; the Alexander von Humboldt Foundation; the Belgian Federal Science Policy Office; the Fonds pour la Formation \`a la Recherche dans l'Industrie et dans l'Agriculture (FRIA-Belgium); the Agentschap voor Innovatie door Wetenschap en Technologie (IWT-Belgium); the Ministry of Education, Youth and Sports (MEYS) of the Czech Republic; the Council of Scientific and Industrial Research, India; the HOMING PLUS programme of the Foundation for Polish Science, cofinanced from European Union, Regional Development Fund, the Mobility Plus programme of the Ministry of Science and Higher Education, the National Science Center (Poland), contracts Harmonia 2014/14/M/ST2/00428, Opus 2014/13/B/ST2/02543, 2014/15/B/ST2/03998, and 2015/19/B/ST2/02861, Sonata-bis 2012/07/E/ST2/01406; the National Priorities Research Program by Qatar National Research Fund; the Programa Clar\'in-COFUND del Principado de Asturias; the Thalis and Aristeia programmes cofinanced by EU-ESF and the Greek NSRF; the Rachadapisek Sompot Fund for Postdoctoral Fellowship, Chulalongkorn University and the Chulalongkorn Academic into Its 2nd Century Project Advancement Project (Thailand); and the Welch Foundation, contract C-1845.

\end{acknowledgments}

\clearpage

\appendix

\section{Supplementary material}
\label{app:yields}

\begin{figure*}[htb]
\centering
\includegraphics[width=1.0\textwidth]{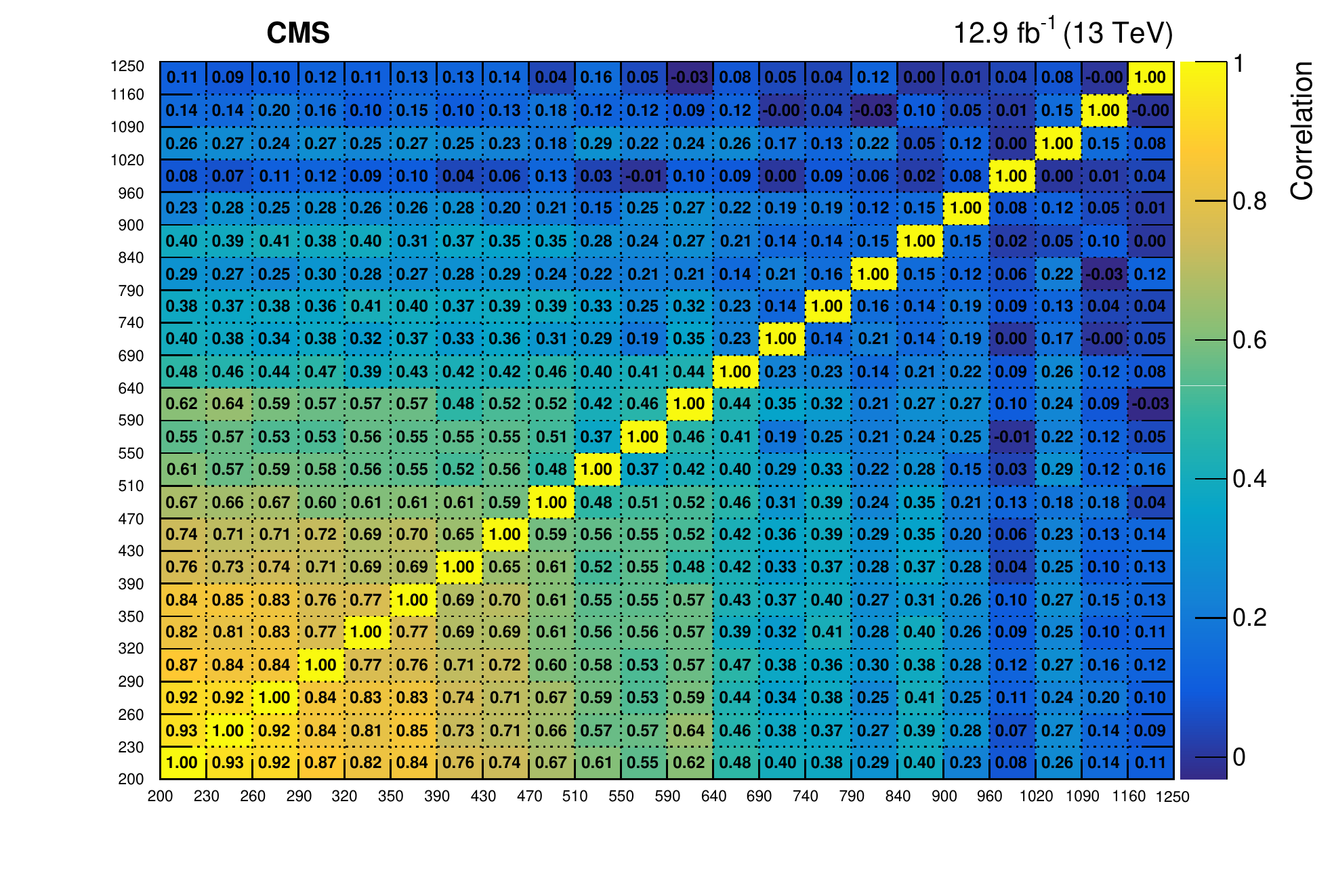}
\caption{Correlations between the uncertainties in the estimated background yields in all the \MET bins of the monojet signal region. The boundaries of the \MET bins, expressed in \GeV,
are shown at the bottom and on the left.}
\label{fig:Correlations_Monojet}

\end{figure*}

\begin{figure*}[htb]
\centering
\includegraphics[width=1.0\textwidth]{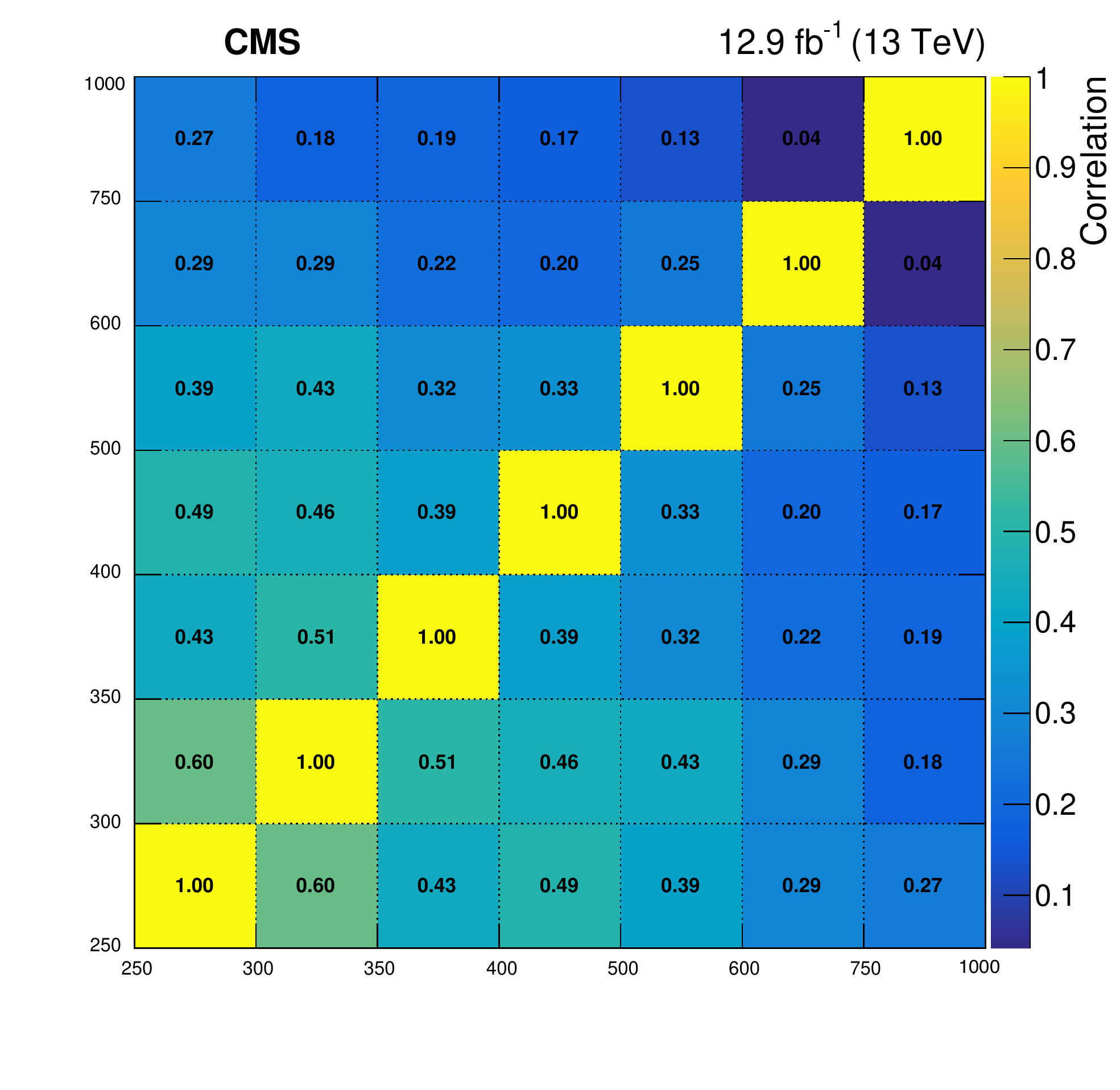}
\caption{Correlations between the uncertainties in the estimated background yields in all the \MET bins of the mono-V signal region. The boundaries of the \MET bins, expressed in \GeV,
are shown at the bottom and on the left.}
\label{fig:Correlations_MonoV}

\end{figure*}

Tables~\ref{tab:MonojetYields} and~\ref{tab:MonoVYields} provide the estimates of various background processes in the monojet and mono-V signal regions, respectively,
that are obtained by performing a fit across all the control samples.
The resulting correlations between the uncertainties in the estimated background yields across all the \MET bins
of the monojet and mono-V signal regions are shown in Fig.~\ref{fig:Correlations_Monojet} and~\ref{fig:Correlations_MonoV}, respectively.

\clearpage

\bibliography{auto_generated}

\cleardoublepage \section{The CMS Collaboration \label{app:collab}}\begin{sloppypar}\hyphenpenalty=5000\widowpenalty=500\clubpenalty=5000\textbf{Yerevan Physics Institute,  Yerevan,  Armenia}\\*[0pt]
A.M.~Sirunyan, A.~Tumasyan
\vskip\cmsinstskip
\textbf{Institut f\"{u}r Hochenergiephysik,  Wien,  Austria}\\*[0pt]
W.~Adam, E.~Asilar, T.~Bergauer, J.~Brandstetter, E.~Brondolin, M.~Dragicevic, J.~Er\"{o}, M.~Flechl, M.~Friedl, R.~Fr\"{u}hwirth\cmsAuthorMark{1}, V.M.~Ghete, C.~Hartl, N.~H\"{o}rmann, J.~Hrubec, M.~Jeitler\cmsAuthorMark{1}, A.~K\"{o}nig, I.~Kr\"{a}tschmer, D.~Liko, T.~Matsushita, I.~Mikulec, D.~Rabady, N.~Rad, B.~Rahbaran, H.~Rohringer, J.~Schieck\cmsAuthorMark{1}, J.~Strauss, W.~Waltenberger, C.-E.~Wulz\cmsAuthorMark{1}
% ...
\vskip\cmsinstskip
% ...
1:~~Also at Vienna University of Technology, Vienna, Austria\\
% ...

\end{sloppypar}
\end{document}